\documentclass[useAMS, usenatbib, final]{iopart}
\expandafter\let\csname equation*\endcsname=\relax 
\expandafter\let\csname endequation*\endcsname=\relax 
\usepackage{amssymb}
\usepackage{amsmath}
\usepackage{array}
\usepackage{graphicx}
\usepackage{color,units}
\usepackage[dvipsnames]{xcolor}
\usepackage{xspace}
\usepackage{dcolumn}
\usepackage{longtable}
\usepackage[normalem]{ulem}
\usepackage{subfigure}
\usepackage[T1]{fontenc}
\usepackage{hyperref}
\graphicspath{ {./} }

\newcommand{\portsmouth}{Institute of Cosmology and Gravitation, University of Portsmouth, Portsmouth, PO1 3FX, UK}

\begin{document}

\title{A rapid multi-modal parameter estimation technique for LISA}

\author{Charlie Hoy$^{1*}$, Connor Weaving$^{1}$, Laura K. Nuttall$^{1}$, Ian Harry$^{1}$}
\address{$^{1}$ \portsmouth}
\address{Email: \mailto{charlie.hoy@port.ac.uk}}
\address{$^{*}$ Author to whom any correspondence should be addressed.}

\begin{abstract}
The Laser Interferometer Space Antenna (LISA) will observe gravitational-wave signals from a wide range of sources, including massive black hole binaries. Although numerous techniques have been developed to perform Bayesian inference for LISA, they are often computationally expensive; analyses often take at least $\sim 1$ month on a single CPU, even when using accelerated techniques. Not only does this make it difficult to concurrently analyse more than one gravitational-wave signal, it also makes it challenging to rapidly produce parameter estimates for possible electromagnetic follow-up campaigns. {\texttt{simple-pe}} was recently developed to produce rapid parameter estimates for gravitational-wave signals observed with ground-based gravitational-wave detectors. In this work, we extend {\texttt{simple-pe}} to produce rapid parameter estimates for LISA sources, including the effects of higher order multipole moments. We show that {\texttt{simple-pe}} infers the source properties of massive black hole binaries in zero-noise at least $\sim 100\times$ faster than existing techniques; $\sim 12$ hours on a single CPU. We further demonstrate that {\texttt{simple-pe}} can be applied before existing Bayesian techniques to mitigate biases in multi-modal parameter estimation analyses of MBHBs.
\end{abstract}

\maketitle

\section{Introduction}

Close to one hundred gravitational-wave (GW) signals have been observed~\cite{LIGOScientific:2016aoc,LIGOScientific:2017vwq,LIGOScientific:2021qlt,KAGRA:2021vkt,Nitz:2021zwj,Olsen:2022pin,Mehta:2023zlk,Wadekar:2023gea} with existing ground-based GW detectors~\cite{LIGOScientific:2014pky,acernese2014advanced,KAGRA:2020tym}. Despite their success, existing and future ground-based GW detectors~\cite{Punturo:2010zz,Reitze:2019iox} are limited to observe GW signals with frequencies in the Hertz to kiloHertz region of the GW spectrium ($> O(\mathrm{Hz})$)~\cite{LIGOScientific:2016wof} owing to various fundamental noise sources on Earth.

The Laser Interferometer Space Antenna (LISA)~\cite{LISA:2017pwj} is a future space-based GW observatory, designed to explore the milliHertz region of the GW spectrum. This frequency range opens up the possibility of observing massive black hole binary (MBHB) coalescences~\cite{Klein:2015hvg}, among other sources~\cite{Glampedakis:2002cb,Babak:2006uv,Babak:2017tow,Nelemans:2001hp}. LISA is expected to observe GW signals produced from MBHB coalescences at large signal-to-noise ratios (SNRs) -- tens to thousands. LISA will therefore provide precise measurements of the properties of MBHBs through Bayesian inference, allowing for stringent tests of general relativity in the strong-field regime, constraining the Hubble constant, and investigating the effect of dark matter on GW propagation~\cite{LISA:2017pwj}.

In order to maximise the scientific return of the LISA mission, it is vital to accurately infer the properties of LISA sources. For MBHBs, this requires the inclusion of higher order multipole moments; a GW signal can be decomposed into a set of spin-weighted spherical harmonics where, for binaries with spins aligned with the orbital angular momentum, the majority of power is contained in the dominant $(\ell, |m|) = (2, 2)$ quadrupole. Additional power may also be radiated in higher order multipole moments, including the $(\ell, |m|) = (3, 3)$ and $(4, 4)$ multipoles~\cite{Thorne:1980ru}. The amplitude of each higher order multipole moment is typically much smaller than the dominant quadrupole, meaning that a direct observation is challenging for low SNR signals; it has been shown previously that only two GW signals to date show clear evidence for higher order multipole moments~\cite{Hoy:2024bbb}. Given that LISA will observe MBHBs at significantly larger SNRs, higher order multipole moments are expected to be observable for most signals~\cite{Mills:2020thr}. It has been shown previously that excluding higher order multipole moments in Bayesian inference analyses can result in biased parameter estimates~\cite{Porter:2008kn,Kumar:2018hml,Kalaghatgi:2019log,Shaik:2019dym,LIGOScientific:2020stg,LIGOScientific:2020zkf,Colleoni:2020tgc,Marsat:2020rtl,Baibhav:2020tma,Katz:2021uax,Krishnendu:2021cyi,Ng:2022vbz,Pratten:2022kug,Pitte:2023ltw,Gong:2023ecg}. 

As we prepare to transition from ground-based to space-based GW data analysis, work needs to be done to significantly reduce the computational cost of performing GW Bayesian inference, particularly when higher order multipole moments are included. It has been shown previously that analyses on mock LISA data complete in $O(10^{5})$ CPU hours compared to $O(10^{3})$ CPU hours for ground-based GW detectors~\cite{Hoy:2024aaa}. Not only does this increased runtime make it difficult to concurrently analyse more than one GW signal, it also makes it challenging to rapidly produce parameter estimates for possible electromagnetic follow-up campaigns. There has been a long history of work that has aimed to reduce the computational cost of GW Bayesian inference~\cite{Canizares:2014fya,Vinciguerra:2017ngf,Wysocki:2019grj,Morisaki:2020oqk,Qi:2020lfr,Morisaki:2021ngj,Williams:2021qyt,Hoy:2022tst,Pathak:2022iar,Rose:2022axr,Dax:2022pxd,Lee:2022jpn,Roulet:2022kot,Yelikar:2023mwg,Morras:2023pug,Pathak:2023ixb,Morisaki:2023kuq,Williams:2023ppp,Wong:2023lgb,Tiwari:2023mzf,Fairhurst:2023idl,Tiwari:2024qzr,Wouters:2024oxj,Vilchez:2024qnw}, and it is natural to investigate these methods for possible LISA applications. One method that has been extensively used in previous work is the heterodyned likelihood~\cite{Cornish:2010kf,Zackay:2018qdy,Cornish:2021lje,Krishna:2023bug}: an optimised likelihood that can reduce the computatioanal cost of LISA analyses by a factor of $10^{2}$ without little loss of accuracy~\cite{Hoy:2024aaa}. Although this significantly reduces the runtime in most cases, analyses of MBHBs with LISA can still take $\sim 1$ month on a single CPU~\cite{Hoy:2024aaa}.

Several software packages have been developed to perform Bayesian inference on LISA sources, including {\texttt{bilby-lisa}}~\cite{Hoy:2024aaa}, {\texttt{pycbc-inference}}~\cite{Weaving:2023fji}, {\texttt{lisabeta}}~\cite{Marsat:2018oam}, {\texttt{glass}}~\cite{Littenberg:2023xpl} and {\texttt{balrog}}~\cite{Pratten:2022kug,Klein:2022rbf}. These software packages make it possible to investigate methods for reducing the computational cost, as well as providing a point of comparison to validate developmental work. We note that several other avenues have been recently explored to reduce the computational cost of Bayesian inference for LISA, including machine learning techniques~\cite{Liang:2024new}. 

In this paper, we outline our developments to {\texttt{simple-pe}}~\cite{Fairhurst:2023idl} for analysing (simulated) LISA data; {\texttt{simple-pe}} is one of the most promising algorithms for reducing the computational-cost of GW Bayesian inference as it provides initial parameter estimates, including the effects of higher order multipole moments, in a matter of CPU minutes for ground-based GW data. We demonstrate that {\texttt{simple-pe}} can be used to infer the initial parameter estimates for simulated MBHB signals observed by LISA in $\sim 12$ hours on a single CPU; $\sim 100\times$ faster than existing techniques that make use of the heterodyned likelihood. We find that the increased computational cost with respect to the ground-based GW counterpart is primarily a result of increased waveform evaluation time. Although {\texttt{simple-pe}} can include any requested (sub)set of higher order multipole moments, we show that over most of the MBHB parameter space, only the $(\ell, |m|) = [(3,3), (4,4)]$ multipoles are required for current MBHB waveform models. However, we predict that future models will need to include the $\ell = m$ multipoles for $\ell > 4$ as they will likely be important for MBHB binaries observed with LISA. We finally discuss possible applications for {\texttt{simple-pe}} beyond providing rapid initial parameter estimates: we show that a natural workflow may produce biased parameter estimates for realistic GW signals, and that {\texttt{simple-pe}} can be easily incorporated into existing workflows to mitigate biases.

The paper is organised as follows: in Section~\ref{sec:simple-pe} we review the {\texttt{simple-pe}} algorithm, and in Section~\ref{sec:modifications} we discuss our modifications to enable rapid parameter estimates for LISA sources; we verify one of the key components of the {\texttt{simple-pe}} algorithm in Section~\ref{sec:verify_metric} and discuss the multipole hierarchy of MBHBs in Section~\ref{sec:mode_higherarchy}. In Section~\ref{sec:zero_noise} we validate our modifications and confirm the effectiveness of {\texttt{simple-pe}} analysing LISA data. Finally in Section~\ref{sec:applications} we discuss possible applications of {\texttt{simple-pe}}, including how it can be used to mitigate biases in multi-modal parameter estimation.

\section{{\texttt{simple-pe}} and modifications for LISA}
\label{sec:overview}

Bayesian inference is the process of estimating the parameters $\boldsymbol{\theta}$of a model $m$ given some observed data $d$. The parameters of the model are represented by the \emph{posterior probability distribution function}, $p(\boldsymbol{\theta} | d, m)$, which is calculated through Bayes' theorem as,

\begin{equation} \label{eq:bayes}
    p(\boldsymbol{\theta} | d, m) = \frac{p(\boldsymbol{\theta} | m)\, p(d | \boldsymbol{\theta}, m)}{\mathcal{Z}},
\end{equation}
where $p(d | \boldsymbol{\theta}, m)$ is the probability of observing the data $d$ given the model parameters $\boldsymbol{\theta}$ and model $m$, otherwise known as the likelihood. $p(\boldsymbol{\theta} | m)$ is the probability that the model has parameters $\boldsymbol{\theta}$ given the model $m$, otherwise known as prior, and $\mathcal{Z}$ is the probability of observing the data given the model $m$, otherwise known as the evidence. For GW astronomy, a Whittle likelihood is often used~\cite{whittle1951hypothesis,Romano:2016dpx,Thrane:2018qnx},

\begin{equation} \label{eq:likelihood}
    p(d | \boldsymbol{\theta}, m) = \exp\left( -\frac{1}{2} (d - m(\boldsymbol{\theta}) | (d - m(\boldsymbol{\theta}))\right),
\end{equation}
where $(a | b)$ denotes the inner product between two frequency series $a(f)$ and $b(f)$,

\begin{equation} \label{eq:inner_product}
    (a | b) = 4 Re\int{df \frac{a(f)b^{*}(f)}{S(f)}},
\end{equation}
where $S(f)$ is the power spectral density (PSD). Under the assumption that the data contains a GW signal $h$, and $h$ is well approximated by $m(\hat{\boldsymbol{\theta}})$, Equation~\ref{eq:likelihood} reduces to,

\begin{equation} \label{eq:reduced_likelihood}
    p(d | \boldsymbol{\theta}, m) \propto \exp\left( - | m(\hat{\boldsymbol{\theta}}) - m(\boldsymbol{\theta})) |^{2} \right),
\end{equation}
where $|a| = \sqrt{(a| a)}$. Often it is convenient to rewrite Equation~\ref{eq:reduced_likelihood} in terms of the match, $\mathfrak{M}$; the match characterises the similarity between two waveforms and ranges between 0 and 1. The match between waveforms $h_{1}$ and $h_{2}$ is defined as,

\begin{equation} \label{eq:match}
    \mathfrak{M} = \max_{dt, d\phi} O(h_{1}, h_{2}) =  \max_{dt, d\phi} \frac{(h_{1} | h_{2})}{|h_{1}| |h_{2}|},
\end{equation}
where $dt$ and $d\phi$ denote the time and phase offset between the two waveforms respectively, and $O(h_{1}, h_{2})$ is known as the overlap. The GW likelihood can then be written in the simplified form,

\begin{equation}\label{eq:likelihood_match}
    p(d | \boldsymbol{\theta}, m) \propto 
    \exp \left( - |m(\hat{\boldsymbol{\theta}})|^{2} \left(1 -  \mathfrak{M}^{2}\right) \right),
\end{equation}
where the peak likelihood is simply,

\begin{equation}
    p(d | \boldsymbol{\theta}, m) \propto 
    \exp \left( - |m(\hat{\boldsymbol{\theta}})|^{2}\right).
\end{equation}

Often it is not possible to trivially evaluate Equation~\ref{eq:bayes}. Consequently, stochastic sampling techniques~\cite{metropolis1949monte,Skilling2004,Skilling:2006gxv} are used to identify the unknown posterior probability distribution function, with numerous packages now available for performing Bayesian inference on GW signals~\cite{Veitch:2014wba,Ashton:2018jfp,Romero-Shaw:2020owr,Biwer:2018osg}, although see e.g.~\cite{Tiwari:2023mzf,Fairhurst:2023idl,Pankow:2015cra,Lange:2018pyp,Delaunoy:2020zcu,Green:2020hst,Chua:2019wwt,Green:2020dnx,Dax:2021tsq,Gabbard:2019rde} for alternative approaches. Unfortunately, performing Bayesian inference on GW signals is often a computationally expensive task. This is because each likelihood calculation involves evaluating the gravitational model $m(\boldsymbol{\theta})$. Similarly, the parameter space is large and highly degenerate. One option to reduce the computational cost is to approximate the likelihood. The heterodyned likelihood~\cite{Cornish:2010kf,Zackay:2018qdy,Cornish:2021lje} was introduced to rapidly estimate the full GW likelihood by assuming that the model $m$ evaluated at two high likelihood points is sufficiently similar, such that the ratio is a smoothly varying function. Summary data for a fiducial point can then be pre-computed, and likelihoods for neighbouring points can be estimated to high accuracy in a fraction of the time. A disadvantage of the heterodyned likelihood is that a point of high likelihood must be known \emph{apriori}. If a fiducial point of low likelihood is chosen, biases may be observed in the inferred posterior distribution. An alternative technique to reduce the computational cost of GW Bayesian inference involves using intuitive arguments to constrain the region of parameter space consistent with the observed data~\cite{Fairhurst:2023idl}.

\subsection{The {\texttt{simple-pe}} algorithm} \label{sec:simple-pe}

{\texttt{simple-pe}} is a parameter estimation algorithm that uses simple, intuitive, and physical insights to estimate the posterior distribution for an observed GW signal. It produces rapid estimates for the model parameters $\boldsymbol{\theta}$ in a matter of CPU minutes for ground-based GW data. At it's core, {\texttt{simple-pe}} relies on the fact that an observation of a chirp-like GW signal provides details about the masses and spins of the binary, while the ability to measure subdominant features in the GW helps to constrain parameter measurements and extrinsic parameters. For instance, it is well known that higher order multipole moments become more pronounced for unequal mass ratio binaries observed edge-on~\cite{Mills:2020thr,Kalaghatgi:2019log}. This means that if significant power beyond the dominant quadrupole is observed in the GW signal, the binary is more likely to have unequal mass components and large inclination angles. Below we provide a brief overview of the {\texttt{simple-pe}} algorithm, but refer the reader to~\cite{Fairhurst:2023idl} for further details.

{\texttt{simple-pe}} constructs a posterior distribution as follows: first, the mass and spin of the best fitting dominant quadrupole-only waveform that maximises the matched filter signal-to-noise ratio (SNR) of the signal, $\hat{\theta}$, is identified. The posterior distribution for the binary's mass and spin is then assumed to be a multi-dimensional Gaussian distribution, where the central value is set to $\hat{\theta}$, and the width is based on the expected accuracy at which the mass and spin can be measured, accounting for known degeneracies throughout the parameter space. SNRs for features beyond the dominant quadrupole are then obtained by matched filtering the data and, based on knowledge of how these SNRs vary as a function of mass, spin and binary orientation, the region of the parameter space consistent with these SNRs is identified. An estimate for the source localisation is then obtained by comparing the arrival times and relative amplitudes/phases of signals observed across the different ground-based GW observatories. Finally a posterior distribution for the luminosity distance of the source is constructed based on the network sensitivity and the inferred uncertainty in the masses and binary orientation.

The posterior distribution constructed by {\texttt{simple-pe}} is reliant on being able to quantify the expected accuracy at which the masses and spins can be measured. {\texttt{simple-pe}} estimates this by calculating the parameter-space metric, $g_{ab}$~\cite{Owen:1995tm}, for an aligned-spin dominant quadrupole only model. The parameter-space metric is defined as,

\begin{equation} \label{eq:metric}
    g_{ab}\delta\boldsymbol{\theta}^{a}\delta\boldsymbol{\theta}^{b} = 1 - \mathfrak{M}
\end{equation}
where $\delta{\boldsymbol{\theta}}$ characterises the parameter separation between two GW signals: $\delta\theta = m(\hat{\boldsymbol{\theta}}) - m(\boldsymbol{\theta})$. Once the metric has been found, an approximate posterior distribution for the masses and spins can be calculated by noting that Equation~\ref{eq:likelihood_match} simplifies to,

\begin{equation}\label{eq:likelihood_metric}
    \begin{split}
        p(d | \boldsymbol{\theta}, m) & \propto 
    \exp \left( - |m(\hat{\boldsymbol{\theta}})|^{2} g_{ab}\delta\boldsymbol{\theta}^{a}\delta\boldsymbol{\theta}^{b} \right) \\
        & \propto 
    \exp \left( - \rho^{2} g_{ab}\delta\boldsymbol{\theta}^{a}\delta\boldsymbol{\theta}^{b} \right), \\
    \end{split}
\end{equation}
where $\rho$ is the observed SNR of the signal. The corresponding 90\% confidence of the posterior distribution can also be estimated through~\cite{Baird:2012cu},

\begin{equation}
    g_{ab}\delta{\boldsymbol{\theta}}^{a}\delta{\boldsymbol{\theta}}^{b} = \frac{3.12}{\rho^{2}}.
\end{equation}
Often the metric is calculated by considering the analytic derivative of the waveform (see e.g.~\cite{Owen:1995tm,Owen:1998dk}), however, {\texttt{simple-pe}} uses finite differencing. This has the advantage that any higher order terms are appropriately included in the metric.

The metric must be calculated to high precision in order to represent the expected accuracy at which the masses and spins can be measured. Unfortunately, the validity of Equation~\ref{eq:metric} is sensitive to the choice of parameters used, with certain configurations giving larger errors than others~\cite{Fairhurst:2023idl,Baird:2012cu,Cutler:1994ys,Ohme:2013nsa,Farr:2014qka}. For this reason {\texttt{simple-pe}} constructs an approximate metric based on the physical mass and spin parameters, and iteratively updates the metric until the principle directions have been found up to a specified tolerance. Samples are then drawn from the 90\% confidence interval, and a posterior is constructed.

As already described, {\texttt{simple-pe}} accounts for higher order multipole content by identifying regions of the parameter space consistent with the observed matched filter SNRs: it computes the higher order multipole content for a GW with parameters $\hat{\theta}$, and matched filters each component to obtain the corresponding SNRs, $\rho_{\ell,m}$, using well-known techniques~\cite{Usman:2015kfa,pycbc-software}. Samples for the extrinsic parameters (including e.g. the binaries inclination angle) are randomly drawn from uniform distributions and, based on the inferred masses and spins calculated from the parameter-space metric, SNRs are calculated for all samples. Samples with SNRs consistent with $\rho_{\ell,m}$ are identified and kept, and the process continues until a pre-determined number of samples are obtained. To reduce the computational cost of calculating $\rho_{\ell,m}$ for all samples, the parameter space is downsampled to $\mathcal{N}$ interpolation points, and dependent variables which are slowly varying are computed ahead of time.

\subsection{Modifications for LISA} \label{sec:modifications}

GW Bayesian analyses are computationally expensive. Unfortunately, the cost increases significantly for LISA applications, with full analyses taking $\sim O(\mathrm{months})$ on a single CPU~\cite{Hoy:2024aaa}. Not only does this make it difficult to concurrently analyse more than one GW signal, it also makes it challenging to rapidly produce and disseminate parameter estimates for possible electromagnetic follow-up campaigns. Given that {\texttt{simple-pe}} has already been shown to produce accurate and reliable posterior distributions in a matter of CPU minutes for ground-based GW observations~\cite{Fairhurst:2023idl}, it is natural to extend this work to produce rapid parameter estimates for LISA signals.

LISA will be composed of three spacecraft, each containing a free falling test mass. Each spacecraft will exchange laser beams, and LISA will observe GW signals by monitoring the distance between free falling test masses as a function of time. Unfortunately, many technical noises need to be suppressed for LISA to achieve the necessary sensitivity to observe GW signals. The most notable is laser frequency noise, which is expected to be several orders of magnitude larger than any change in arm length due to expected GW signals~\cite{LISA:2017pwj}. Time Delay Interferometry (TDI) is a post-processing technique that creates a combination of virtual interferometers which a) suppresses laser frequency noise by 8 orders of magnitude, and b) remains sensitive to GW signals. Although numerous TDI combinations have been introduced, second-generation TDI allows for linearly changing arm lengths making them the most applicable for LISA applications~\cite{Tinto:1999yr,armstrong1999time,Estabrook:2000ef,Vallisneri:2005ji,Tinto:2020fcc,tinto2023second}. Unfortunately, second-generation TDI contains correlated noise properties. An additional linear transformation is therefore often applied to form a combination of virtual interferometers with uncorrelated noise properties: the $A$, $E$ and $T$ channels~\cite{Prince:2002hp}. Similar to previous work~\cite{Hoy:2024aaa,Weaving:2023fji}, we have added LISA to {\texttt{simple-pe}} by treating the observatory as a network of distinct virtual interferometers, each corresponding to separate TDI observables. Although in this work we restrict attention to only a single TDI channel for simplicity (the $A$ TDI channel), it is straightforward to extend the algorithm to include data from the other virtual interferometers.

As described in Section~\ref{sec:simple-pe}, a GW model is needed to a) identify the best fitting dominant quadrupole-only waveform, b) calculate the parameter-space metric, and c) obtain the matched filter SNRs in higher order multipole moments. Given that {\texttt{simple-pe}} treats LISA as a network of virtual interferometers, it is natural to build a waveform generator that returns the expected GW as seen in each TDI channel. In this work we have extended the {\texttt{simple-pe}} algorithm to interface with the {\texttt{BBHx}} library~\cite{Katz:2021uax,Katz:2020hku,michael_katz_2021_5730688}. {\texttt{BBHx}} is an open source software package that generates GW polarizations, and returns the expected GWs in the $A$, $E$ and $T$ TDI channels. Importantly for this work, {\texttt{BBHx}} interfaces with {\texttt{IMRPhenomHM}} and therefore is able to return not only the dominant quadrupole, but also specific higher order multipole moments; {\texttt{IMRPhenomHM}} is a model that includes the $(\ell, m) = [(2, 2), (2, 1), (3, 3), (3, 2), (4, 4), (4, 3)]$ multipoles, and restricts the black hole spins to be aligned with the total orbital angular momentum~\cite{London:2017bcn}. {\texttt{BBHx}} is able to reduce the computational cost of generating waveforms by using interpolation: a reduced set of frequencies are used to evaluate the waveform and interpolation is used to provide the waveform at the desired frequency points~\cite{Katz:2021uax,Katz:2020hku,michael_katz_2021_5730688}. Since {\texttt{simple-pe}} is designed to produce rapid estimates for the posterior distribution, we enable the interpolation in {\texttt{BBHx}} by default. We note that {\texttt{BBHx}} accounts for the orbit of LISA around the Sun via the time and frequency dependent transformation $\mathcal{T}(f, t)$~\cite{Marsat:2020rtl}. 

As described, {\texttt{simple-pe}} infers the source localisation by comparing the arrival times and relative amplitudes/phases of signals observed across different GW observatories. For LISA this is challenging since it requires knowing the exact position of all spacecraft as a function of time. An additional complication is that the sky location is highly degenerate for LISA: for aligned-spin dominant quadrupole only models, we expect to infer one of eight sky locations. The inclusion of higher order multipoles generally improves the source localization, but a bimodality is still expected~\cite{Marsat:2020rtl}. In this work, we have implemented an alternative method to estimate the source localisation in {\texttt{simple-pe}}, which takes advantage of the degeneracies in~\cite{Marsat:2020rtl} and builds upon the idea of sky folding introduced in~\cite{Weaving:2023fji}. We split the sky into eight octants and uniformly draw samples for a single case. The SNR in the dominant quadrupole is calculated for all samples, using the inferred masses and spins calculated from the parameter-space metric, and samples with SNRs consistent with the SNR in the dominant quadrupole $\rho_{22}$ are identified and kept. The other seven degenerate sky locations are then reconstructed using known transformations in~\cite{Marsat:2020rtl}. Once the source localisation consistent with the dominant quadrupole has been identified, we infer which of the eight degenerate sky locations are consistent with the observed matched filter SNRs in the higher order multipoles. 

\subsubsection{Verifying the parameter-space metric for LISA} \label{sec:verify_metric}

\begin{figure}
    \centering
    \includegraphics[width=0.7\textwidth]{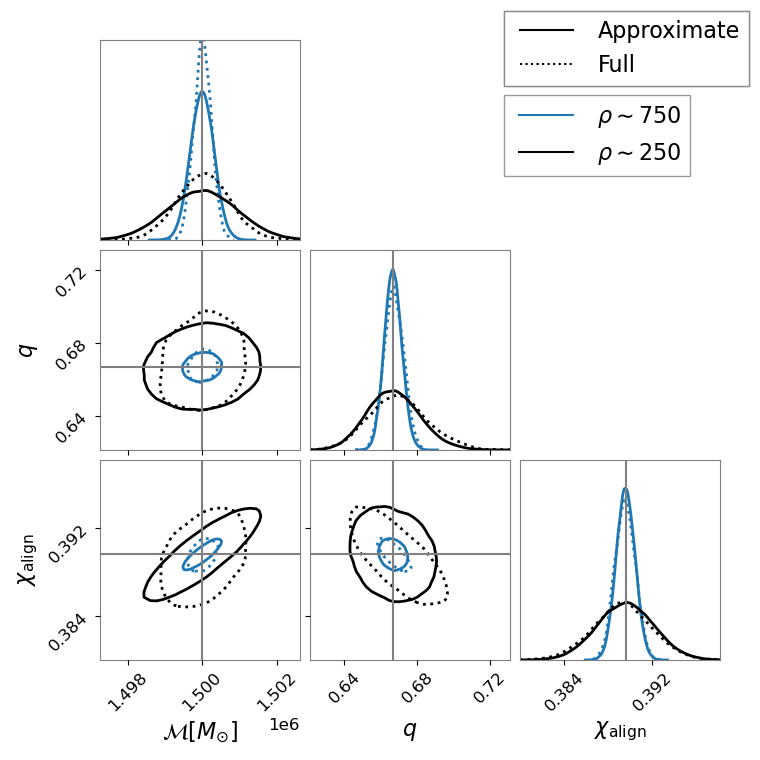}
    \caption{Corner plot showing the posterior distribution for the chirp mass $\mathcal{M}$, mass ratio $q = m_{2} / m_{1}$, and effective aligned spin $\chi_{\mathrm{align}}$ (see Equation~\ref{eq:align}) when analysing a GW produced by a MBHB with parameters: $\mathcal{M} = 1.5\times 10^{6}\, M_{\odot}$, $q = m_{2} / m_{1} = 0.67$, $\chi_{1} = 0.5$, $\chi_{2} = 0.2$, $\beta = 0.73$, $\lambda = 5.9$ and $\theta_{JN} = \pi / 3\, \mathrm{rad}$ (see text for parameter definitions). We consider two signal-to-noise ratio simulations, $\rho\sim 250$ in black and $\rho\sim 750$ in blue. The \emph{Approximate} posterior distribution based on the parameter-space metric is shown by the solid lines, and the \emph{Full} posterior distribution obtained by performing full Bayesian inference is shown by the dotted lines. The grey cross hairs show the true values, and the 2-dimensional contours show the inferred 90\% confidence interval. Both analyses used only the A TDI channel, and all parameters other than the chirp mass, mass ratio and aligned spin were fixed to their true values.}
    \label{fig:metric_corner}
\end{figure}

The parameter-space metric is vital in {\texttt{simple-pe}} as it is directly used for estimating the masses and spins, and indirectly used for constraining the sky location of the source. Here, we verify that a) our modifications to {\texttt{simple-pe}} allow us to accurately infer the parameter-space metric, and b) allow us to reconstruct posteriors for the masses and spins of MBHBs consistent with full Bayesian inference analyses\footnote{We expect the parameter-space metric to remain accurate for LISA applications since it can be derived from fisher matrix approximations which have been shown to work well for LISA sources~\cite{Vecchio:2003tn,Berti:2004bd,Lang:2006bsg,Arun:2008zn}.}.

We simulate a GW produced by a MBHB with chirp mass $\mathcal{M} = (m_{1} m_{2})^{3/5} / (m_{1} + m_{2})^{1/5} = 1.5\times 10^{6}\, M_{\odot}$, mass ratio $q = m_{2} / m_{1} = 0.67$, primary spin $\chi_{1} = 0.5$ and secondary spin $\chi_{2} = 0.2$ at two distinct SNRs $\rho\sim 250$ and $\rho\sim 750$. We assume the MBHB has spins aligned with the total orbital angular momentum, and located at an ecliptic latitude $\beta = 0.73$ and ecliptic longitude $\lambda = 5.9$, viewed at an inclination angle $\theta_{JN} = \pi / 3\, \mathrm{rad}$. The phase and polarization were set to be $\phi = 3.8$ and $\psi = 4.0$ respectively and the injection is defined in the LISA reference frame. We scale the luminosity distance of the source until we achieve the requested SNR. We inject at two SNRs to test our algorithm for both highly peaked likelihood surfaces (high SNR) and more correlated parameter spaces (low SNR). To verify the results from {\texttt{simple-pe}}, we performed Bayesian inference with {\texttt{bilby-lisa}}~\cite{Hoy:2024aaa} over four dimensions -- the chirp mass $\mathcal{M}$, mass ratio $q$ and the two aligned spin components $\chi_{1}, \chi_{2}$ -- with the {\texttt{dynesty}} nested sampler~\cite{Speagle:2019ivv}. We keep all other parameters fixed to their true values and use the full Whittle likelihood. All analyses excluded higher order multipole moments\footnote{{\texttt{simple-pe}} constructs the parameter-space metric for dominant quadrupole-only models, as described in Section~\ref{sec:simple-pe}.}. We set $\hat{\theta}$ to the injected values when generating the parameter-space metric, but we note that {\texttt{simple-pe}} has methods to identify the mass and spin of the best fitting dominant quadrupole-only waveform that maximises the matched filter SNR of the signal~\cite{Fairhurst:2023idl}.

In Figure~\ref{fig:metric_corner} we compare the posterior distributions obtained with {\texttt{simple-pe}} and {\texttt{bilby-lisa}}. To remain consistent with \cite{Fairhurst:2023idl}, we show the posterior distribution for $\chi_{\mathrm{align}}$, an alternative effective spin parameter that better describes the degeneracy in individual component spins, defined as,

\begin{equation} \label{eq:align}
    \chi_{\mathrm{align}} = \frac{m_{1}^{\alpha}\chi_{1} + m_{2}^{\alpha}\chi_{2}}{m_{1}^{\alpha} + m_{2}^{\alpha}}
\end{equation}
where $\alpha = 4/3$. We see that in general {\texttt{simple-pe}} accurately estimates the one- and two-dimensional marginalized posterior distributions, with 90\% confidence intervals consistent with {\texttt{bilby-lisa}} for both SNRs. We see that {\texttt{simple-pe}} is able to recover the degeneracies in this reduced parameter space, but it slightly over estimates them for $\mathcal{M} - \chi_{\mathrm{align}}$, and slightly underestimates them for $q - \chi_{\mathrm{align}}$ degeneracies, as is seen more clearly in the $\rho\sim 250$ injection. In general, we see that {\texttt{simple-pe}} performs better at larger SNRs, which is expected as the likelihood surface is more gaussian, inline with the underlying assumption from the metric approximation.

\subsubsection{Multipole hierarchy for MBHBs}
\label{sec:mode_higherarchy}

It is well known that omitting higher order multipole moments in Bayesian analyses can bias the inferred properties from GW observations, with numerous studies demonstrating this for ground-based GW observations~\cite{Kumar:2018hml,Kalaghatgi:2019log,Shaik:2019dym,LIGOScientific:2020stg,LIGOScientific:2020zkf,Colleoni:2020tgc,Krishnendu:2021cyi,Ng:2022vbz}, as well as anticipated LISA sources~\cite{Porter:2008kn,Marsat:2020rtl,Baibhav:2020tma,Katz:2021uax,Pratten:2022kug,Pitte:2023ltw,Gong:2023ecg}. {\texttt{simple-pe}} accounts for the effects of higher order multipole moments by identifying the region of the parameter space consistent with the observed SNR in each higher order multipole.

\begin{figure}
    \centering
    \includegraphics[width=0.7\textwidth]{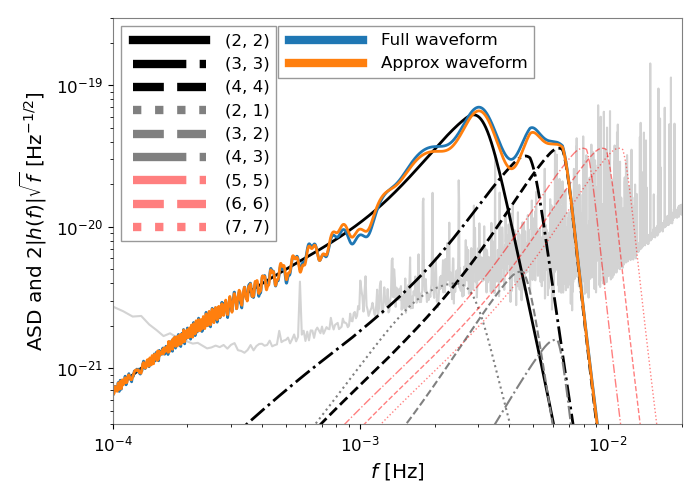}
    \caption{Frequency evolution of the gravitational-wave produced by a MBHB with parameters: $\mathcal{M} = 2.0\times 10^{6}\, M_{\odot}$, $q = m_{2} / m_{1} = 0.5$, $\chi_{1} = -0.5$, $\chi_{2} = -0.1$, $\beta = 1.3$, $\lambda = 4.8$ and $\theta_{JN} = \pi / 3\, \mathrm{rad}$ (see text for parameter definitions). We show the Full waveform, which was produced with the {\texttt{IMRPhenomHM}} waveform model as implemented in {\texttt{BBHx}}, as well as the individual $(\ell, |m|) = [(2, 2), (2, 1), (3, 3), (3, 2), (4, 4), (4, 3)]$ multipoles. We also show predictions for the $(\ell, |m|) = [(5, 5), (6, 6), (7, 7)]$ multipoles, based on multiplying the frequency evolution of the $(4, 4)$ by $5/4$, $6/4$ and $7/4$ respectively. The Approx waveform is simply the sum of the leading contributions to the waveform: the $(\ell, |m|) = [(2, 2), (3, 3), (4, 4)]$ multipoles. For simplicity, we only show the multipoles projected in the A TDI channel and for comparison, we show a representative noise curve for the A TDI channel in LISA. The noise curve was calculated from the Sangria data challenge and includes galactic white dwarf binaries. Each waveform and multipole is plotted such that the area between each line and the noise curve is proportional to the SNR.}
    \label{fig:mode_higherarchy}
\end{figure}

For ground-based GW observations, {\texttt{simple-pe}} restricts attention to only the $(\ell, m) = [(2, 2), (3, 3)]$ multipoles. This is because additional higher order multipole moments will have a negligible impact on the observed GW for typical stellar mass binary black holes observed with e.g. LIGO, Virgo, KAGRA~\cite{LIGOScientific:2014pky,acernese2014advanced,KAGRA:2020tym,Mills:2020thr}. However, additional higher order multipoles may become significant for e.g. MBHBs where the total mass and SNRs are expected to be significantly larger~\cite{Mills:2020thr}.

Previous work has investigated the importance of higher order multipoles for MBHBs observed with LISA~\cite{Pitte:2023ltw,Gong:2023ecg}, but there is disagreement concerning which multipoles are needed to avoid biases in parameter estimation analyses. Although {\texttt{simple-pe}} can include any requested set of higher order multipole moments\footnote{Although {\texttt{simple-pe}} in principle can include any requested set of higher order multipole moments, we are restricted by the higher order multipole moments available in the GW model.}, there is an advantage to using a subset to reduce computational cost. Here, we revisit the multipole hierarchy of MBHBs, and identify which multipoles contribute most significantly for MBHBs. Rather than deciphering this through detailed parameter estimation analyses as has been done previously, we instead compare the frequency evolution of the individual higher order multipole components against a representative noise curve for LISA; if a significant fraction of a given multipole lies beneath the noise curve, it will likely have a small SNR and can safely be neglected.

\begin{figure*}
    \centering
    \includegraphics[width=1.0\textwidth]{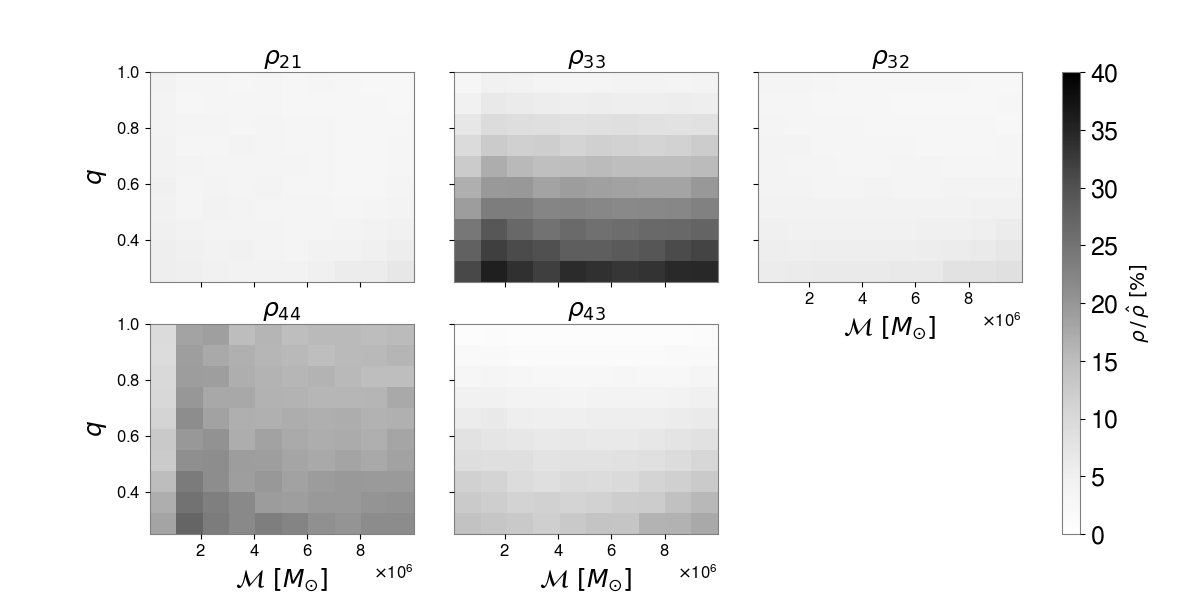}
    \caption{Plot showing how the orthogonal SNR, $\rho$, of the $(\ell, |m|) = [(2, 2), (2, 1), (3, 3), (3, 2), (4, 4), (4, 3)]$ multipoles vary with respect to the total SNR of the GW signal, $\hat{\rho}$, across the MBHB parameter space. We calculated the orthogonal SNR for $35,000$ randomly chosen MBHBs, and projected the results onto the chirp mass $\mathcal{M}$ and mass ratio $q$ parameter space. Darker regions indicate a larger contribution to the total SNR of the GW signal. For simplicity, we only considered the A TDI channel.}
    \label{fig:mode_snr}
\end{figure*}

Figure~\ref{fig:mode_higherarchy} shows the frequency evolution of a simulated GW for a MBHB with parameters: $\mathcal{M} = 2.0\times 10^{6}\, M_{\odot}$, $q = m_{2} / m_{1} = 0.5$, $\chi_{1} = -0.5$, $\chi_{2} = -0.1$, $\beta = 1.3$, $\lambda = 4.8$ and $\theta_{JN} = \pi / 3\, \mathrm{rad}$. The waveforms are normalized such that the area between each signal and the noise curve is proportional to the SNR, see Equation 1 and surrounding text in~\cite{LIGOScientific:2016dsl} for details. For this signal, we see that the $(2, 2)$ multipole dominates the GW emission, as expected, extending from $\sim 2\times 10^{-4} - 3\times 10^{-2}\,\mathrm{Hz}$. Although the $(3, 3)$ and $(4, 4)$ multipoles have a lower amplitude than the $(2, 2)$, they extend to higher frequencies~\cite{Leaver:1985ax,Berti:2005ys,Roy:2019phx} and consequently contribute to the total SNR of this signal. As expected~\cite{Blanchet:2013haa}, the $\ell \neq m$ multipoles ($(2, 1)$, $(3, 2)$ and $(4, 3)$) have amplitudes significantly below the $(3, 3)$ and $(4, 4)$, implying that they have negligable impact on the observed waveform. Indeed, we see that a waveform constructed from just the $[(2, 2), (3, 3), (4, 4)]$ multipoles approximates the full waveform to high accuracy, with a match between waveforms $\mathfrak{M} = 0.998$.

Since GWs can be written as an infinite sum of higher order multipole moments, an interesting question is whether additional multipoles beyond $\ell = 4$ significantly contribute to the total SNR of the signal. Unfortunately, since {\texttt{IMRPhenomHM}} includes only a subset of these higher order multipole moments, we are unable to accurately produce the frequency evolution of e.g. the $(\ell, m) = (5, 5)$ multipole\footnote{We note that recently GW models have started to implement multipoles beyond $\ell = 4$, with \cite{Varma:2018mmi,Cotesta:2018fcv,Pompili:2023tna} including the $(5, 5)$ multipole, but they are not included within the {\texttt{BBHx}} library.}. However, estimates of the $\ell = m$ multipoles beyond $\ell = 4$ can be made by multiplying the frequency evolution of the $(4, 4)$ projected in the A TDI channel by $\ell / 4$, while assuming the same amplitude. This estimate assumes that the frequency evolution of higher order multipoles is $\ell / 2$ times the frequency evolution of the $(2, 2)$~\cite{Leaver:1985ax,Berti:2005ys,Roy:2019phx}, as well as noting that all higher order multipoles become more significant with higher masses~\cite{Mills:2020thr}. We note that while the amplitudes of the $\ell > 4$ higher order multipole moments are well known from post Newtonian theory~\cite{Blanchet:2013haa}, we are unable to make use of this since their mapping to the A TDI channel is non-trivial. For instance, the frequency evolution of each higher order multipole moment in LISA is,

\begin{equation} \label{eq:hm_waveform}
    h_{\ell, m}^{A}(f, t_{\ell, m}(f)) = \mathcal{T}^{A}(f, t_{\ell, m}) h_{\ell, m}(f)
\end{equation}
where $\mathcal{T}(f, t_{\ell, m})$ is the time and frequency dependent transformation function, $h_{\ell, m}(f)$ is the frequency evolution of the $(\ell, m)$ multipole and,

\begin{equation}
    t_{\ell, m} = t_{c} - \frac{1}{2\pi}\frac{d\phi_{\ell, m}}{df}
\end{equation}
where $t_{c}$ is the coalescence time of the binary and $\phi_{\ell, m}$ is the phase of the $(\ell, m)$ multipole. Consequently, we need to know the phase derivative of each higher order multipole moment in order to calculate $\mathcal{T}(f, t_{\ell, m})$, and hence estimate the amplitude as seen in LISA. 

As shown in Figure~\ref{fig:mode_higherarchy}, we expect the $\ell > 4$ multipoles to actively contribute to the total SNR of this signal. We also expect that this will become more apparent for MBHBs with larger mass. This is because as the total mass of the signal increases, the merger frequency of the GW signal decreases, and multipoles beyond $\ell = 6$ spend longer within the sensitive region of LISA. We anticipate that there will be high mass systems where the $(\ell, |m|) = (2, 2)$ quadrupole lies outside of the sensitive region of LISA while higher order multipole moments will be observable in the GW signal\footnote{\cite{Fairhurst:2023beb} made a similar argument for next generation ground-based GW detectors.}. We therefore expect that neglecting multipoles beyond $\ell = 6$ will significantly bias our parameter estimation. Based on the binary considered, we estimate that this occurs for masses $M \gtrsim 7.5\times 10^{8}\, M_{\odot}$. Of course, this is not specific to LISA and we expect a similar phenomenon to occur with current ground-based GW detectors.

So far we have only considered a single MBHB and we expect the relative importance of each higher order multipole to vary for different MBHB configurations (see e.g. Figure 3 in \cite{Pitte:2023ltw}). We therefore next investigate the importance of each multipole across the MBHB parameter space. Rather than plotting and comparing each multipole with a representative noise curve, we calculate and compare SNRs. We randomly draw $\sim 35,000$ MBHBs, with chirp masses $10^{5}\, M_{\odot} < \mathcal{M} < 10^{7}\, M_{\odot}$, mass ratios $0.25 < q < 1.0$ and spins $0 < \chi < 0.99$. We fix the luminosity distance to be $50 \mathrm{Gpc}$ for all cases, and all other parameters are randomly chosen from an unconstrained boundary. For simplicity we only consider the $A$ TDI channel, and restrict attention to MBHBs with total SNR greater than 10, which is the adopted threshold for MBHB detection~\cite{LISA:2017pwj}.

When calculating the SNR in each higher order multipole, we take care to only consider the component orthogonal to the dominant $(2, 2)$ quadrupole. This is because only the orthogonal component is observable, with any parallel contribution acting only to increase the observed SNR in the $(2, 2)$ quadrupole~\cite{Mills:2020thr}\footnote{Although the orthogonal SNR is only defined for ground-based GW detectors~\cite{Mills:2020thr}, we apply it for LISA signals in this work. We argue that this approach is valid for MBHBs since their GW signals rapidly accumulate SNR within the final day of merger~\cite{Marsat:2020rtl}, meaning that LISA can be assumed approximately stationary during this signal}. The SNR orthogonal to the dominant quadrupole is simply,

\begin{equation} \label{eq:orthogonal_snr}
    \rho_{\ell, m} = |h_{\ell, m}(f)| \sqrt{1 - O(h_{\ell, m}, h_{2,2})^{2}},
\end{equation}
where $O(h_{\ell, m}, h_{2,2})$ is the overlap between $h_{\ell, m}$ and $h_{2, 2}$ as defined in Equation~\ref{eq:match}, and the orthogonal component of each higher order multipole is, 

\begin{equation}
    h_{\ell, m}^{\perp} = h_{\ell, m} - \frac{(h_{2,2} | h_{\ell, m})}{|h_{2, 2}|^{2}} h_{2, 2}.
\end{equation}

As shown in Figure~\ref{fig:mode_snr}, the least significant multipoles over the MBHB parameter space considered are the $(2, 1)$ and $(3, 2)$ and $(4, 3)$ harmonics, with each multipole contributing on average $\sim 4\%$, $4\%$ and $7\%$ of the total SNR of the signal respectively. There are cases where the $(2, 1)$, $(3, 2)$ multipoles contribute more than $10\%$ of the total SNR, but this is only for around $5\%$ of MBHBs considered. This increases to around $25\%$ for the $(4, 3)$ multipole. We find that in general the $(2, 1)$, $(3, 2)$ and $(4, 3)$ multipoles become more significant for MBHBs that are observed edge-on, with spins anti-aligned with the orbital angular momentum. For the $(3, 2)$ and $(4, 3)$ multipoles in particular, the relative importance becomes more apparent for small mass ratio systems with high chirp masses. Interestingly, we find that the $(4, 3)$ and $(3, 2)$ multipoles are on average more significant than the $(2, 1)$. At first this seems to be in contention with the results presented in \cite{Pitte:2023ltw}, however we emphasise that unlike \cite{Pitte:2023ltw}, we are considering the orthogonal components of each multipole, and therefore only the observable component.

We see that the $(3, 3)$ and $(4, 4)$ higher order multipoles are the most significant across the MBHB parameter space considered, with each contributing on average $18\%$ of the total SNR of the signal. As expected, the $(3, 3)$ becomes more significant for smaller mass ratio systems observed edge-on (i.e. at $\theta_{JN} = \pi / 2$), and in general, is the most significant higher order multipole. On the other hand, we find that the $(4, 4)$ does not vary significantly with mass ratio, but remains significant for edge-on systems. Interestingly, the $(4, 4)$ can on average be more significant than the $(3, 3)$ for MBHBs with mass ratios $> 0.9$. If we assume that a higher order multipole is observable for SNRs $> 2.1$~\cite{Mills:2020thr}, we find that the $(3, 3)$ multipole will nearly always be observable: the orthogonal SNR is greater than $2.1$ in $99\%$ of binaries considered.

In light of these results, we expect that the $[(2, 2), (3, 3), (4, 4)]$ higher order multipoles to be dominant for MBHBs (when restricting attention to $\ell \leq 4$), with other multipoles likely to have a negligable impact on the observed GW signal. For this reason we restrict {\texttt{simple-pe}} to use only the $[(2, 2), (3, 3), (4, 4)]$ multipoles in this work. However, we note that it may be necessary to include the $\ell = m$ multipoles for $\ell > 4$ when waveform models are developed which include greater higher order multipole content.

\section{Simulated signal recovery in zero-noise} \label{sec:zero_noise}

To demonstrate the efficacy of {\texttt{simple-pe}} for rapid parameter estimation with LISA, we analyse a MBHB signal and compare results with full Bayesian inference analyses. We simulate a GW signal for a MBHB with parameters: $\mathcal{M} = 1.5\times 10^{6}\, M_{\odot}$, $q = m_{2} / m_{1} = 0.67$, $\chi_{1} = 0.5$, $\chi_{2} = 0.2$, $\beta = 0.73$, $\lambda = 5.9$, $\theta_{JN} = \pi / 3\, \mathrm{rad}$ and $d_{L} = 35\, \mathrm{Gpc}$ defined in the LISA reference frame, with the {\texttt{IMRPhenomHM}} waveform model (including all available multipoles) as implemented in {\texttt{BBHx}}. For simplicity, we restrict attention to only the $A$ TDI channel, and inject this signal at a SNR $\rho\sim 470$; the SNR in the dominant quadrupole $\rho_{22} \sim 420$, the $(3, 3)$ multipole $\rho_{33} \sim 100$ and the $(4, 4)$ multipole $\rho_{44} \sim 160$. We chose to inject the signal at an SNR between the two cases studied in Section~\ref{sec:verify_metric} to verify that {\texttt{simple-pe}} can handle more correlated parameter spaces while still ensuring a realistic SNR for observed MBHB signals with LISA. Given that the true parameters are known, biases in the inferred posterior distributions can be identified.

\begin{figure*}
    \centering
    \includegraphics[width=0.7\textwidth]{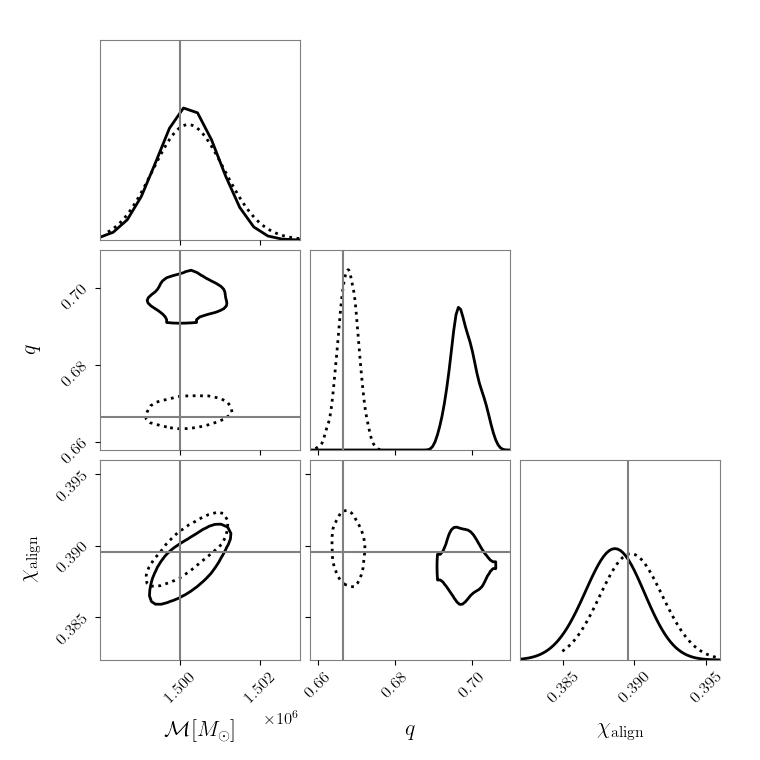}
    \caption{Corner plot showing the posterior distribution for the inferred chirp mass $\mathcal{M}$, mass ratio $q = m_{2} / m_{1}$, and aligned spin $\chi_{\mathrm{align}}$ for a MBHB injected into zero-noise with parameters: $\mathcal{M} = 1.5\times 10^{6}\, M_{\odot}$, $q = m_{2} / m_{1} = 0.67$, $\chi_{1} = 0.5$, $\chi_{2} = 0.2$, $\beta = 0.73$, $\lambda = 5.9$, $\theta_{JN} = \pi / 3\, \mathrm{rad}$ and $d_{L} = 35\, \mathrm{Gpc}$. The \emph{Approximate} posterior distribution shown by the solid black lines is based on our modified {\texttt{simple-pe}} algorithm, and the \emph{Full} posterior distribution shown by the dotted black lines is obtained by performing full Bayesian inference. The grey cross hairs show the true values, and the 2-dimensional contours show the inferred 90\% confidence interval. Both analyses used only the A TDI channel.}
    \label{fig:zero_noise_intrinsic}
\end{figure*}

\begin{figure*}
    \centering
    \includegraphics[width=0.7\textwidth]{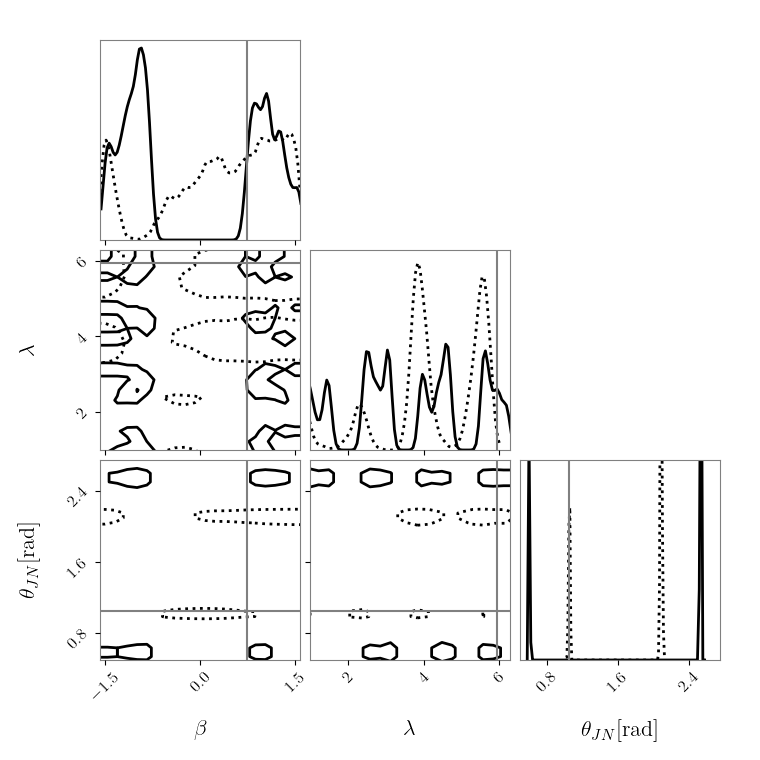}
    \caption{Corner plots showing the posterior distribution for the inferred ecliptic latitude $\beta$, ecliptic longitude $\lambda$ and inclination angle $\theta_{JN}$ for a MBHB injected into zero-noise with parameters given in Figure~\ref{fig:zero_noise_intrinsic}. As in Figure~\ref{fig:zero_noise_intrinsic}, the \emph{Approximate} posterior distribution shown by the solid black lines is based on our modified {\texttt{simple-pe}} algorithm, and the \emph{Full} posterior distribution shown by the dotted black lines is obtained by performing full Bayesian inference. The grey cross hairs show the true values, and the 2-dimensional contours show the inferred 90\% confidence interval. Both analyses used only the A TDI channel.}
    \label{fig:zero_noise_extrinsic}
\end{figure*}

As in Section~\ref{sec:verify_metric}, we compare results with those obtained by {\texttt{bilby-lisa}} and the full Whittle likelihood. We take advantage of the {\texttt{dynesty}} nested sampler with 1000 live points and the {\texttt{bilby}} implemented {\texttt{rwalk}} sampling algorithm with an average of 60 accepted steps per MCMC. We integrated the likelihood between $10^{-4} < f < 10^{-1}\, \mathrm{Hz}$ with the {\texttt{IMRPhenomHM}} waveform model as implemented in {\texttt{BBHx}}. We include all multipoles available in {\texttt{IMRPhenomHM}} in order to verify the the conclusions in Section~\ref{sec:mode_higherarchy}. We sampled in the LISA reference frame to remain consistent with the injection, and we used uninformative and wide priors for all parameters. We sampled in the standard parameters: chirp mass, mass ratio, aligned spin components, merger time, distance, inclination $\theta_{JN}$, ecliptic longitude $\delta$ and latitude $\alpha$, phase and polarization. We also used a PSD that included galactic white dwarf binary confusion noise. As in Section~\ref{sec:verify_metric} we set $\hat{\theta}$ to the injected values when generating the parameter-space metric in {\texttt{simple-pe}}. However, we note that this is not an idealised case as {\texttt{simple-pe}} would likely identify the injected values from any starting point owing to the inbuilt maximisation routines~\cite{Fairhurst:2023idl}. 

In Figures~\ref{fig:zero_noise_intrinsic} and ~\ref{fig:zero_noise_extrinsic} we see that in general, the injected values lie within the 90\% confidence intervals inferred by {\texttt{simple-pe}} for most parameters. When considering the sky location, we see that {\texttt{simple-pe}} is unable to break the well known sky location degeneracy as discussed in Section~\ref{sec:modifications} and~\cite{Marsat:2020rtl}. We suspect that this is because there is insufficient power in higher order multipoles to reduce the exact eight-fold symmetry to the expected bimodal sky position. Interestingly, we see more support in the ecliptic latitude at the true value, while the ecliptic longitude remains fairly unconstrained.

We see a small bias in the inferred mass ratio and inclination angle measurements with the injected value lying outside of the 90\% confidence interval; we infer a mass ratio that is too large, and an inclination angle that is too small. Although the mass ratio and inclination angle measurements are biased, it is unlikely to affect the astrophysical interpretation of the source.

\begin{figure}
    \centering
    \includegraphics[width=0.7\textwidth]{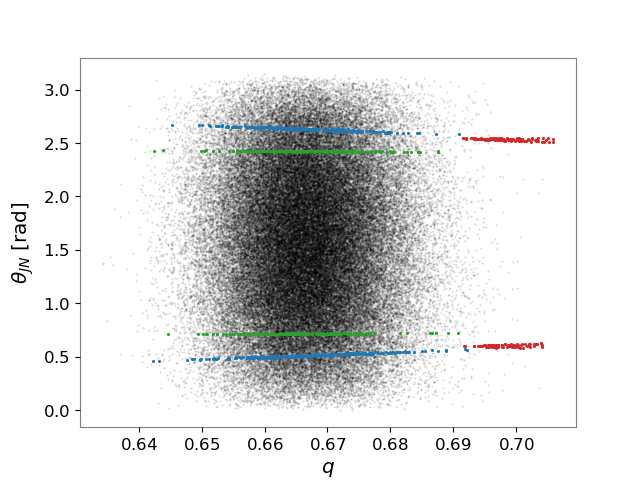}
    \caption{Scatter plot showing the randomly chosen samples (black) that are consistent with the observed matched filter SNRs in the $(\ell, m) = (3, 3)$ (blue), $(4, 4)$ (green) and the $[(3, 3), (4, 4)]$ multipoles (red) across the mass ratio $q$ and inclination angle $\theta_{JN}$ parameter space. Each sample is drawn from the parameter-space metric and assigned a randomly chosen chosen set of extrinsic parameters.}
    \label{fig:snr_parameter_space_points}
\end{figure}

\begin{figure}
    \centering
    \includegraphics[width=0.7\textwidth]{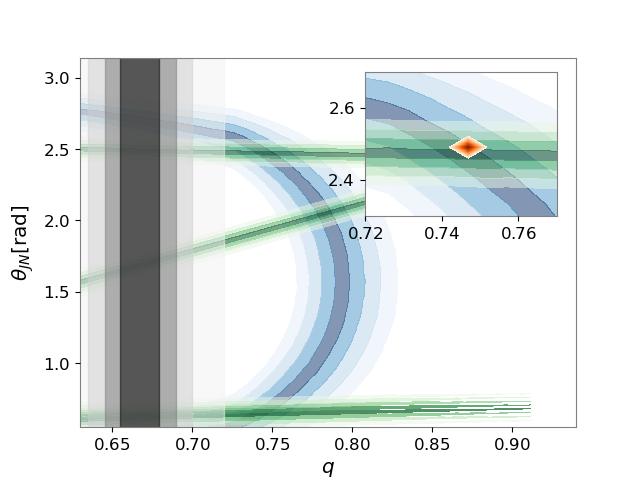}
    \caption{Distribution of $\rho_{3,3}$ (Blue) and $\rho_{4,4}$ (Green) across the mass ratio $q$ and inclination angle $\theta_{JN}$ parameter space for a MBHB injected into zero-noise with parameters given in Figure~\ref{fig:zero_noise_intrinsic}. Constraints on the mass ratio from the parameter-space metric are shown in Grey. Darker colours show regions of higher probability based upon the observed matched filter SNRs in the $(\ell, m) = (3, 3)$ and $(4, 4)$ multipoles. The inset shows a zoom-in on one of the three areas consistent with both the observed matched filter SNRs in the $(\ell, m) = (3, 3)$ and $(4, 4)$ multipoles, and the orange contours show the combined probability.}
    \label{fig:snr_parameter_space}
\end{figure}

To investigate why {\texttt{simple-pe}} infers a biased mass ratio and inclination measurement, in Figure~\ref{fig:snr_parameter_space_points} we plot the samples drawn from the parameter-space metric that are compatible with the observed SNRs in the $(3, 3)$ and $(4, 4)$ multipoles. We see that although the majority of mass ratio samples are centered around the true value, the only samples that are consistent with the observed SNRs in the $(3, 3)$ and $(4, 4)$ multipoles are for $0.69 < q < 0.71$. This implies that although a large number of samples are randomly drawn, only a few are kept, decreasing the efficiency of the sampler. In Figure~\ref{fig:snr_parameter_space} we plot $\rho_{\ell, m}$ across the mass ratio and inclination angle parameter space, while keeping all other parameters fixed to their true values. We see that there are three areas consistent with both observed SNRs in the $(3, 3)$ and $(4, 4)$ multipoles. We see that the inferred match filter SNRs rule out inclination angles $0.6 < \theta_{\mathrm{JN}} < 2.0$ and mass ratios $q < 0.7$ and $q > 0.8$. Figures~\ref{fig:snr_parameter_space_points} and~\ref{fig:snr_parameter_space} imply that we see a biased mass ratio and inclination angle measurement because the injected values lie within a region inconsistent with the observed SNRs. This hints at an inconsistency during the initial matched filtering step where the observed SNRs in the higher order multipoles are determined. If we increase the observed SNRs in the $(3, 3)$ and $(4, 4)$ multipoles by 20\%, {\texttt{simple-pe}} recovers a mass ratio and inclination posterior peaking at the true values. We rule out any issues arising from using an interpolant (to rapidly calculate the slowly varying dependent variables, see Section~\ref{sec:simple-pe} for details) since the distribution of $|h_{\ell, m}(f)|$ calculated directly with {\texttt{BBHx}} is comparable across the same parameter space.

When comparing with the results to a full Bayesian analysis we see that in general {\texttt{simple-pe}} and {\texttt{bilby-lisa}} obtain comparable posteriors. Interestingly {\texttt{bilby-lisa}} was also unable to break the well known sky location degeneracy as discussed in Section~\ref{sec:modifications} and~\cite{Marsat:2020rtl}, although there is a mild preference for the true ecliptic latitude and longitude. This corroborates the idea that there is insufficient power in higher order multipoles to break this degeneracy. As expected, {\texttt{bilby-lisa}} obtained a more accurate mass ratio and inclination measurement, with the inferred posteriors peaking at the injected values. Given that {\texttt{simple-pe}} only uses information from the $(\ell, m) = [(2, 2), (3, 3), (4, 4)]$ multipoles to constrain the parameter space, while {\texttt{bilby-lisa}} includes all multipoles available in {\texttt{IMRPhenomHM}}, we see that for this case additional higher order multipoles moments are not needed to infer the posterior distribution. This supports the conclusions found in Section~\ref{sec:mode_higherarchy}. 

{\texttt{simple-pe}} completed in $\sim 12$ CPU hours while {\texttt{bilby-lisa}} with the full whittle likelihood took $\sim 10^{5}$ CPU hours; a $> 10^{4}\times$ reduction in the CPU time. Of course, if the accelerated heterodyned likelihood was used rather than the full Whittle likelihood, we expect a full Bayesian analysis to complete $\sim 10^{2}\times$ faster~\cite{Hoy:2024aaa}; indeed although not presented here, an analysis with {\texttt{pycbc-inference}} and the heterodyned likelihood took $\sim 10^{3}$ CPU hours\footnote{Unlike {\texttt{bilby-lisa}}, the heterodyned likelihood in {\texttt{pycbc-inference}} is valid for higher order multipole waveform models. {\texttt{pycbc-inference}} evaluates the heterodyned likelihood for waveform models that include higher order multipole moments by splitting the waveform model and evaluating the heterodyned likelihood for each multipole separately, while taking care to account for cross-terms~\cite{Leslie:2021ssu,Weaving:2024the}.}. We note that this difference in timing is based on a single case, and we expect the timing to change depending on the MBHB GW signal, especially the injected SNR and SNRs in higher order multipole moments. When comparing different Bayesian methods, it is can also be useful to compare e.g. the number of likelihood evaluations in order to gauge the efficiency of the algorithms. Unfortunately, this cannot be done with {\texttt{simple-pe}} since the likelihood is not directly computed. 

\section{Applications of {\texttt{simple-pe}}} \label{sec:applications}

As discussed throughout this paper, the heterodyned likelihood is often used in full Bayesian analyses as it has the potential to reduce the computational cost by a factor of $10^{2}$~\cite{Hoy:2024aaa}. However, in order to take advantage of this, a fiducial point of high likelihood must be known \emph{apriori}. A natural approach is to set the fiducial point to be the best fitting template from a matched filter search~\cite{Krishna:2023bug,Weaving:2023fji}; a matched filter search identifies possible GW candidates by matching the observed data against a pre-generated bank of GW templates. A bank is typically constructed through stochastic placement~\cite{Harry:2009ea, Ajith:2012mn}, and is considered converged when any randomly generated GW signal within the target parameter space is covered by at least one template, up to a minimum match $\mathfrak{M}$.

\cite{Weaving:2023fji} recently adapted {\texttt{pycbc}}~\cite{Usman:2015kfa,pycbc-software} to perform a matched filter search for MBHBs with LISA. They generated a bank of fifty dominant quadrupole-only templates with minimum match $\mathfrak{M} = 0.9$, and they demonstrated that MBHBs could not only be identified in the mock LISA data, but the best fitting template could also be used as the fiducial point for the heterodyned likelihood. \cite{Weaving:2023fji} found their template bank was sufficient in obtaining fiducial points that did not bias Bayesian analyses of aligned-spin dominant quadropole-only injections.

The matched filter search introduced in~\cite{Weaving:2023fji} is limited to dominant quadrupole-only waveform models, inline with existing pipelines for ground-based GW observatories~\cite{Nitz:2021zwj, Mehta:2023zlk,Sachdev:2019vvd, Andres:2021vew}. Matched filter searches have traditionally neglected higher order multipole content to simplify the search, reduce the parameter space, and decrease the computational cost of the analysis (although see e.g.~\cite{Harry:2017weg,Chandra:2022ixv,Wadekar:2023kym} for alternative approaches). Although dominant quadrupole-only searches will likely not affect the ability to observe the majority of MBHB signals with LISA, it may impact downstream Bayesian analyses since the best matching template will likely no longer be adequate for use with the heterodyned likelihood.

To investigate possible biases from using a dominant quadrupole-only template with the heterodyned likelihood for GW signals, we simulated an end-to-end pipeline: we injected a MBHB GW signal that includes higher order multipole content into mock LISA data, identified the best matching dominant quadrupole-only template by performing the matched filter search described in~\cite{Weaving:2023fji}, and recovered the properties using {\texttt{pycbc-inference}} with the fiducial point set to the best fitting template from the search. We used {\texttt{pycbc-inference}} for the Bayesian analysis to remain consistent with the search.

We obtained mock LISA data by subtracting the MBHB GW signals from the Sangria training set~\cite{le_jeune_2022_7132178}. After subtracting the simulated MBHB GW signals, the Sangria training set contains Gaussian instrumental noise and simulated GWs from 30 million Galactic white dwarf binaries, and 17 verification Galactic binaries. We then injected a GW signal produced by a MBHB with chirp mass $\mathcal{M} \sim 9.8\times 10^{5} \mathrm{M_{\odot}}$, mass ratio $q \sim 0.96$ and spins $-0.5$ and $-0.45$ for the primary and secondary black hole respectively\footnote{The injection was chosen to have parameters equal to the second injection in the Sangria blind data challenge~\cite{le_jeune_2022_7132178}. However, unlike the signals in Sangria blind data challenge itself, we included higher order multipole moments by interfacing the {\texttt{IMRPhenomHM}} waveform model.}, as defined in the solar system baricenter (SSB) reference frame. We used the {\texttt{IMRPhenomHM}} implementation in the {\texttt{BBHx}} package to generate the simulated GW, and restricted the signal to include only the dominant $(\ell, |m|) = [(2, 2), (3, 3), (4, 4)]$ multipoles to reduce the computational cost (see Section~\ref{sec:mode_higherarchy} for details). We used 3000 live points to improve convergence and the {\texttt{dynesty}} implemented {\texttt{rwalk}} sampling algorithm. All others settings are consistent with those outlined in Section~\ref{sec:zero_noise}. To reduce the computational cost of the Bayesian analysis, we fixed the phase to that found by the matched filter search.

\begin{figure*}
    \centering
    \includegraphics[width=0.98\textwidth]{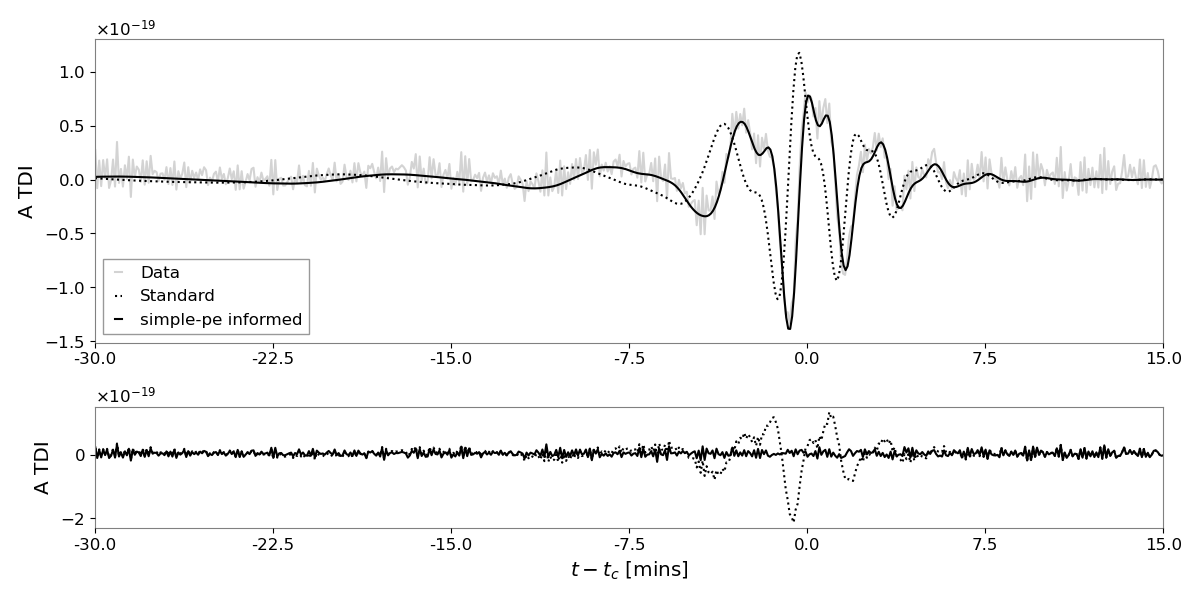}
    \caption{\emph{Top}: Comparison between the reconstructed GW signal obtained when using the results from {\texttt{simple-pe}} to inform the fiducial point for the heterodyned likelihood (black solid), the reconstructed GW signal obtained when using the best fitting template from a matched filter search for the heterodyned likelihood (black dot) and the time-domain mock LISA data which includes a simulated MBHB GW signal (grey). The reconstructed GW signal was obtained by analysing the A TDI channel and using the {\texttt{IMRPhenomHM}} waveform model as implemented in {\texttt{BBHx}}. We shift the reconstructed GW signal and data by the inferred merger time as defined in the LISA frame $t_{c}$. The reconstructed GW signal is plotted as a band representing the 90\% credible region. \emph{Bottom}: Comparison between the residuals after subtracting each reconstructed GW signals from the time-domain mock LISA data.}
    \label{fig:simple_pe_waveform_comparison}
\end{figure*}

The search outlined in \cite{Weaving:2023fji} found a best fitting template with parameters: $\mathcal{M} = 9.3\times 10^{5}\, M_{\odot}$, $q = m_{2} / m_{1} = 0.9$, $\chi_{1} = -0.5$, $\chi_{2} = -0.1$, $\beta = 1.3$, $\lambda = 4.8$, $\theta_{JN} = 2.4\, \mathrm{rad}$, $\psi=3.0$. When using this as our fiducial point for the heterodyned likelihood, we observed a biased posterior distribution. This was expected since the fiducial point used in the heterodyned technique was not of sufficiently high likelihood; indeed the match between the best fitting template and the simulated signal was $0.88$\footnote{Although the bank was constructed minimum match $\mathfrak{M} = 0.9$, this does not necessary mean that any randomly generated GW signal that includes the effects of higher order multipole moments will be covered by at least one template, up to a minimum match $\mathfrak{M}$. This is because the bank was constructed to cover GW signals including only the dominant quadrupole. This means that all randomly generated GW signals that include only the dominant quadrupole within the target parameter space will be covered by at least one template, up to a minimum match $\mathfrak{M}$, but not GW signals including higher order multipole moments.}. We find that all marginalized posterior distributions were biased, with none incorporating the injected value. Interestingly, we found that the posterior did shift away from the fiducial point but not enough to include the injected values; for instance the inferred chirp mass shifted from a fiducial value of $9.3\times 10^{5}\, M_{\odot}$ to an inferred $9.793^{+0.003}_{-0.004} \times 10^{5}\, M_{\odot}$. We also found no evidence for bimodalities in the inferred posterior. In fact, the only concerning artefact was the apparent `railing' in the spin of the larger black hole against extremal spin -- we inferred a primary spin of $-0.988^{+0.008}_{-0.002}$. However, we note `railing' in the inferred spin posterior has been observed with current GW observations at significantly lower SNRs (see e.g.~\cite{Hannam:2021pit}).

Our analysis resulted in a reconstructed GW signal that does not resemble the data, see Figure~\ref{fig:simple_pe_waveform_comparison}. In fact, when subtracting the reconstructed GW signal from the data, we obtain a residual that is different from the expectation from Gaussian noise coloured by the PSD. Given that the marginalized posterior distributions are all Gaussian with no signs of poorly converged results, there would be no reason to cast doubt on the inferred posterior. This highlights the importance of calculating and looking at the residuals. Since the residual contains power beyond the expectation from Gaussian noise, we argue that this analysis would not be suitable for a global fit style approach (see e.g.~\cite{Littenberg:2023xpl}) since biased results and/or additional non-existent signals in the data may be observed.

There are several approaches to mitigate this biased result. First, a template bank that includes higher order multipole moments can be developed. For this case the best fitting template is guaranteed to be of high likelihood, assuming that the bank is sufficiently large to fully cover the target parameter space, and the minimum match is appropriately chosen. As already discussed higher order multipole moments have so far been neglected for ground-based GW searches and, given that LISA data analysis is generally more computationally expensive than it's ground based GW counter-part, we argue that this is currently unrealistic (although see~\cite{Wadekar:2023kym} which may provide methods for implementing an efficient higher order multipole search in the future). Second, a dominant quadrupole only template bank can be constructed with an improved minimum match threshold. While this is possible and will provide a higher likelihood fiducial point, the computational cost of generating a bank with e.g. $\mathfrak{M} = 0.95$ increases by a factor of $\sim 40$ compared to a bank with $\mathfrak{M} = 0.9$; a bank with $\mathfrak{M} = 0.9$ takes $\sim 32$ CPU hours to converge. Finally, we can refine the best fitting template from a dominant quadrupole only search by using rapid multi-modal parameter estimation techniques to find a fiducial point of higher likelihood. We argue that running a rapid multi-modal parameter estimation technique is preferred since although a bank is typically constructed prior to an observing period, a rapid parameter estimation technique will be cheaper if $\lesssim 100$ MBHBs are observed during LISA's lifetime\footnote{Theoretical models predict different estimates for the merger rate of MBHBs. As a result, the current best guess for the expected number of MBHBs observed with LISA is anywhere between $O(1)$ to $O(100)$ per year~\cite{Sesana:2021jfh}}; this is assuming that {\texttt{simple-pe}} takes on average $12$ CPU hours per analysis. 

\subsection{Mitigating biases by employing {\texttt{simple-pe}}}

The idea of refining an initial point of low likelihood for use with the heterodyned technique is not new: \cite{Katz:2021uax} found that refinement was needed to perform Bayesian inference with just the dominant quadrupole, and {\texttt{bilby-lisa}} includes methods to refine an initial point through a pre-computed optimisation routine (through the standard {\texttt{bilby}} infrastructure~\cite{Krishna:2023bug}). Specifically \cite{Katz:2021uax} proposed updating the fiducial point every 5000 iterations of the sampler. However, the idea of using a rapid parameter estimation algorithm to refine an initial point to include the effects of higher order multipole moments is novel, and we argue significantly faster and more reliable than using previous optimisations.

To investigate how {\texttt{simple-pe}} can be used to mitigate biases in multi-modal Bayesian inference analyses, we simulated an end-to-end pipeline: we injected the same MBHB signal as in Section~\ref{sec:applications} and similarly identified the best matching dominant quadrupole-only template by performing the matched filter search described in~\cite{Weaving:2023fji} via {\texttt{pycbc}}. We then performed {\texttt{simple-pe}} to obtain initial parameter estimates, and recovered the properties using {\texttt{pycbc-inference}} with the fiducial point set to the highest SNR sample found by {\texttt{simple-pe}}. Of course, we could have designed a pipeline that avoided running the matched filter search all together since {\texttt{simple-pe}} has inbuilt methods to identify the best fitting dominant quadrupole-only template, as discussed in Secs.~\ref{sec:verify_metric} and~\ref{sec:zero_noise}. However, since a search may be run as part of a LISA pipeline, it is natural to use the results from a matched filter search to seed {\texttt{simple-pe}}. We note that one difference between the results presented here and those presented in Section~\ref{sec:applications} is that for this case, the phase was fixed to the injected value while previously it was set to the output from the search. This is because {\texttt{simple-pe}} does not currently sample over the phase but we intend to include this parameter in future work.

For this case {\texttt{simple-pe}} completed in a fraction of the time compared to the previous example (see Section~\ref{sec:zero_noise} for details). The reason is because the observed SNRs in the $(3, 3)$ and $(4, 4)$ multipoles were consistent with the majority of points drawn from the parameter-space metric. When {\texttt{simple-pe}} is used to set the fiducial point for the heterodyned likelihood, we see remarkable agreement between the reconstructed GW obtained and the data, see Figure~\ref{fig:simple_pe_waveform_comparison}. In fact, when subtracting the reconstructed GW from the data, the residual is comparable with expectations from Gaussian noise coloured by the PSD, implying that the MBHB signal has been appropriately subtracted from the data. When inspecting the marginalized posterior distributions we see that the injected values are within the 90\% confidence interval for all parameters except for the polarization, where we injected $\psi=0.4$ and recovered $\psi = 2.3^{+0.5}_{-0.2}$.

Interestingly, we found that when {\texttt{simple-pe}} is used to set the fiducial point for the heterodyned likelihood, the analysis completed $1.3\times$ faster compared to when the best fitting template from the search is used: $\sim 870$ CPU hours compared to $\sim 1100$ CPU hours. This correlates with a reduced number of likelihood evaluations, \texttt{ncall} as reported by {\texttt{dynesty}}, implying a greater sampling efficiency: $6\times 10^{6}$ compared to $7.9\times 10^{6}$. However, we note that we are only considering a single case in this work, and therefore we cannot guarantee that this is a general result across the MBHB parameter space; further work would need to be done on numerous injections at different SNRs.

\section{Discussion}

Inferring the properties of LISA sources is crucial to maximise the scientific return of the LISA mission. Although numerous algorithms have been developed to perform Bayesian inference on LISA data in recent years, most share a common trait: they are computationally expensive, even when using algorithms to optimise the likelihood computation. Not only does this make it difficult to concurrently analyse more than one GW signal, it also makes it challenging to rapidly produce parameter estimates for possible electromagnetic follow-up campaigns.

In this work we introduce our modifications to {\texttt{simple-pe}}, a parameter estimation toolkit that produces rapid parameter estimates, including the effects of higher order multipole moments, to analyse (simulated) LISA data. Although there has been a long history of work that has aimed to reduce the computational cost of GW Bayesian inference, with numerous algorithms developed, we modify {\texttt{simple-pe}} as it uses simple, intuitive and physical insights to estimate posterior distributions for GW signals in a matter of CPU minutes for existing GW detectors. It has also been shown to produce reliable results for GW signals with strong imprints of higher order multipole moments, a region of the parameter space where GW models are often expensive to generate, and the parameter space is often harder to sample effectively with conventional methods.

By focusing on MBHBs, we demonstrate that the foundations of the {\texttt{simple-pe}} algorithm can be used on LISA data; specifically we show that the parameter-space metric, used to quantify the expected accuracy at which parameters can be measured, gives consistent results with those obtained when performing full Bayesian inference. We then discuss the multipole hierarchy of MBHBs and show that for current waveform models the $(\ell, |m|) = [(2, 2), (3, 3), (4, 4)]$ multipoles dominate, and other multipole moments can often be neglected. However, we predict that this will need to revised to include the $\ell = m$ multipoles for $\ell > 4$ when waveform models are developed which include greater higher order multipole content. As a result, we restrict {\texttt{simple-pe}} to use only the dominant $(\ell, |m|) = [(2, 2), (3, 3), (4, 4)]$ multipole moments. We then show that {\texttt{simple-pe}} is able to infer the properties of MBHBs in zero-noise in $\sim 12$ hours on a single CPU; a $> 100\times$ reduction in the CPU time compared with full Bayesian inference analyses using optimised likelihoods. We highlight that {\texttt{simple-pe}} is not optimised, and the CPU time can likely be reduced by profiling the code and reducing bottlenecks. We note that this timing comparison is based on a single analysis, and we expect the exact timings to change depending on the simulation, especially the injected SNR and the SNR in the higher order multipole moments. We also emphasise that, in its current state, {\texttt{simple-pe}} \emph{will not} always provide the most accurate parameter estimates, but it will massively reduce the parameter space volume consistent with the observed signal.

We finally discuss a \emph{realistic} application for {\texttt{simple-pe}}: we show that a natural workflow where the best fitting template from a dominant quadrupole-only search is used as the fiducial point for the heterodyned likelihood will produce biased results for realistic MBHB GW signals for LISA. We then show that these biases can be mitigated by refining the fiducial point based on the results from {\texttt{simple-pe}}. This has applications in possible global fit approaches (see e.g.~\cite{Littenberg:2023xpl,Katz:2024oqg}), as well for more general population studies of MBHBs. For instance, it has been shown previously that the computational cost of a global fit pipeline can be reduced by including a pre-optimisation routine~\cite{Strub:2024kbe}. Although we have demonstrated that {\texttt{simple-pe}} can be used to mitigate biases for LISA specific simulations, a similar approach may also be applicable for ground-based GW data analysis pipelines. We note that in principle this application is not specific to the {\texttt{simple-pe}} algorithm, and any rapid multi-modal parameter estimation technique for LISA will likely obtain similar conclusions.

Although the results presented in this paper have focused specifically on MBHBs and a single TDI channel, a future extension would be to include waveform models from other sources~\cite{LISAConsortiumWaveformWorkingGroup:2023arg} as well as additional TDI channels. Although not discussed in this work, {\texttt{simple-pe}} for ground-based GW detectors also includes the effects of the general relativistic phenomenon of spin-induced orbital precession~\cite{Apostolatos:1994mx}, and has been shown to produce reliable results for binary systems with significant in-plane spins. Given the promising results that we have presented for multi-modal parameter estimation with {\texttt{simple-pe}}, an interesting extension would be to investigate the impact of spin-precession. Similarly, as we have maintained the general structure of {\texttt{simple-pe}} in our modifications, additional physics can be incorporated in {\texttt{simple-pe}} to obtain more accurate and reliable parameter estimation results.

\section*{Acknowledgements}

We would like to thank Stephen Fairhurst for comments on this manuscript as well as valuable discussions throughout this project. We are also grateful to Mark Hannam for conversations about the hierarchy of higher order multipoles, Michael Williams for help debugging {\texttt{pycbc-inference}}, and Jonathan Thompson for general discussions throughout this project. CH, CW and LN thank the UKRI Future Leaders Fellowship for support through the grant MR/T01881X/1. IH thanks the STFC for support through the grants ST/T000333/1 and ST/V005715/1. This work used the computational resources provided by the ICG, SEPNet and the University of Portsmouth, supported by STFC grant ST/N000064, and the DiRAC@Durham facility managed by the Institute for Computational Cosmology on behalf of the STFC DiRAC HPC Facility (www.dirac.ac.uk). The equipment for the latter was funded by BEIS capital funding via STFC capital grants ST/P002293/1, ST/R002371/1 and ST/S002502/1, Durham University and STFC operations grant ST/R000832/1. DiRAC is part of the National e-Infrastructure. This work used \texttt{numpy}~\cite{harris2020array}, \texttt{scipy}~\cite{2020SciPy-NMeth}, \texttt{pesummary}~\cite{Hoy:2020vys}, \texttt{pycbc}~\cite{pycbc-software} and \texttt{matplotlib}~\cite{2007CSE.....9...90H}.

\section*{References}
\bibliographystyle{unsrt}
\bibliography{main}

\begin{thebibliography}{100}

\bibitem{LIGOScientific:2016aoc}
B.~P. Abbott et~al.
\newblock {Observation of Gravitational Waves from a Binary Black Hole Merger}.
\newblock {\em Phys. Rev. Lett.}, 116(6):061102, 2016.

\bibitem{LIGOScientific:2017vwq}
B.~P. Abbott et~al.
\newblock {GW170817: Observation of Gravitational Waves from a Binary Neutron
  Star Inspiral}.
\newblock {\em Phys. Rev. Lett.}, 119(16):161101, 2017.

\bibitem{LIGOScientific:2021qlt}
R.~Abbott et~al.
\newblock {Observation of Gravitational Waves from Two Neutron
  Star\textendash{}Black Hole Coalescences}.
\newblock {\em Astrophys. J. Lett.}, 915(1):L5, 2021.

\bibitem{KAGRA:2021vkt}
R.~Abbott et~al.
\newblock {GWTC-3: Compact Binary Coalescences Observed by LIGO and Virgo
  during the Second Part of the Third Observing Run}.
\newblock {\em Phys. Rev. X}, 13(4):041039, 2023.

\bibitem{Nitz:2021zwj}
Alexander~H. Nitz, Sumit Kumar, Yi-Fan Wang, Shilpa Kastha, Shichao Wu, Marlin
  Sch\"afer, Rahul Dhurkunde, and Collin~D. Capano.
\newblock {4-OGC: Catalog of Gravitational Waves from Compact Binary Mergers}.
\newblock {\em Astrophys. J.}, 946(2):59, 2023.

\bibitem{Olsen:2022pin}
Seth Olsen, Tejaswi Venumadhav, Jonathan Mushkin, Javier Roulet, Barak Zackay,
  and Matias Zaldarriaga.
\newblock {New binary black hole mergers in the LIGO-Virgo O3a data}.
\newblock {\em Phys. Rev. D}, 106(4):043009, 2022.

\bibitem{Mehta:2023zlk}
Ajit~Kumar Mehta, Seth Olsen, Digvijay Wadekar, Javier Roulet, Tejaswi
  Venumadhav, Jonathan Mushkin, Barak Zackay, and Matias Zaldarriaga.
\newblock {New binary black hole mergers in the LIGO-Virgo O3b data}.
\newblock {\em arXiv e-prints}, page arXiv:2311.06061, 11 2023.

\bibitem{Wadekar:2023gea}
Digvijay Wadekar, Javier Roulet, Tejaswi Venumadhav, Ajit~Kumar Mehta, Barak
  Zackay, Jonathan Mushkin, Seth Olsen, and Matias Zaldarriaga.
\newblock {New black hole mergers in the LIGO-Virgo O3 data from a
  gravitational wave search including higher-order harmonics}.
\newblock {\em arXiv e-prints}, page arXiv:2312.06631, 12 2023.

\bibitem{LIGOScientific:2014pky}
J.~Aasi et~al.
\newblock {Advanced LIGO}.
\newblock {\em Class. Quant. Grav.}, 32:074001, 2015.

\bibitem{acernese2014advanced}
F~Acernese, M~Agathos, K~Agatsuma, D~Aisa, N~Allemandou, A~Allocca, J~Amarni,
  P~Astone, G~Balestri, G~Ballardin, et~al.
\newblock Advanced virgo: a second-generation interferometric gravitational
  wave detector.
\newblock {\em Classical and Quantum Gravity}, 32(2):024001, 2014.

\bibitem{KAGRA:2020tym}
T.~Akutsu et~al.
\newblock {Overview of KAGRA: Detector design and construction history}.
\newblock {\em PTEP}, 2021(5):05A101, 2021.

\bibitem{Punturo:2010zz}
M.~Punturo et~al.
\newblock {The Einstein Telescope: A third-generation gravitational wave
  observatory}.
\newblock {\em Class. Quant. Grav.}, 27:194002, 2010.

\bibitem{Reitze:2019iox}
David Reitze et~al.
\newblock {Cosmic Explorer: The U.S. Contribution to Gravitational-Wave
  Astronomy beyond LIGO}.
\newblock {\em Bull. Am. Astron. Soc.}, 51(7):035, 2019.

\bibitem{LIGOScientific:2016wof}
Benjamin~P Abbott et~al.
\newblock {Exploring the Sensitivity of Next Generation Gravitational Wave
  Detectors}.
\newblock {\em Class. Quant. Grav.}, 34(4):044001, 2017.

\bibitem{LISA:2017pwj}
Pau Amaro-Seoane et~al.
\newblock {Laser Interferometer Space Antenna}.
\newblock {\em arXiv e-prints}, page arXiv:1702.00786, 2 2017.

\bibitem{Klein:2015hvg}
Antoine Klein et~al.
\newblock {Science with the space-based interferometer eLISA: Supermassive
  black hole binaries}.
\newblock {\em Phys. Rev. D}, 93(2):024003, 2016.

\bibitem{Glampedakis:2002cb}
Kostas Glampedakis, Scott~A. Hughes, and Daniel Kennefick.
\newblock {Approximating the inspiral of test bodies into Kerr black holes}.
\newblock {\em Phys. Rev. D}, 66:064005, 2002.

\bibitem{Babak:2006uv}
Stanislav Babak, Hua Fang, Jonathan~R. Gair, Kostas Glampedakis, and Scott~A.
  Hughes.
\newblock {'Kludge' gravitational waveforms for a test-body orbiting a Kerr
  black hole}.
\newblock {\em Phys. Rev. D}, 75:024005, 2007.
\newblock [Erratum: Phys.Rev.D 77, 04990 (2008)].

\bibitem{Babak:2017tow}
Stanislav Babak, Jonathan Gair, Alberto Sesana, Enrico Barausse, Carlos~F.
  Sopuerta, Christopher P.~L. Berry, Emanuele Berti, Pau Amaro-Seoane, Antoine
  Petiteau, and Antoine Klein.
\newblock {Science with the space-based interferometer LISA. V: Extreme
  mass-ratio inspirals}.
\newblock {\em Phys. Rev. D}, 95(10):103012, 2017.

\bibitem{Nelemans:2001hp}
G.~Nelemans, L.~R. Yungelson, and Simon~F. Portegies~Zwart.
\newblock {The gravitational wave signal from the galactic disk population of
  binaries containing two compact objects}.
\newblock {\em Astron. Astrophys.}, 375:890--898, 2001.

\bibitem{Thorne:1980ru}
K.~S. Thorne.
\newblock {Multipole Expansions of Gravitational Radiation}.
\newblock {\em Rev. Mod. Phys.}, 52:299--339, 1980.

\bibitem{Hoy:2024bbb}
Charlie Hoy, Stephen Fairhurst, and Ilya Mandel.
\newblock {Precession and higher order multipoles in binary black holes (and
  lack thereof)}.
\newblock {\em arXiv e-prints}, page arXiv:2408.03410, 8 2024.

\bibitem{Mills:2020thr}
Cameron Mills and Stephen Fairhurst.
\newblock {Measuring gravitational-wave higher-order multipoles}.
\newblock {\em Phys. Rev. D}, 103(2):024042, 2021.

\bibitem{Porter:2008kn}
Edward~K. Porter and Neil~J. Cornish.
\newblock {The Effect of Higher Harmonic Corrections on the Detection of
  massive black hole binaries with LISA}.
\newblock {\em Phys. Rev. D}, 78:064005, 2008.

\bibitem{Kumar:2018hml}
Prayush Kumar, Jonathan Blackman, Scott~E. Field, Mark Scheel, Chad~R. Galley,
  Michael Boyle, Lawrence~E. Kidder, Harald~P. Pfeiffer, Bela Szilagyi, and
  Saul~A. Teukolsky.
\newblock {Constraining the parameters of GW150914 and GW170104 with numerical
  relativity surrogates}.
\newblock {\em Phys. Rev. D}, 99(12):124005, 2019.

\bibitem{Kalaghatgi:2019log}
Chinmay Kalaghatgi, Mark Hannam, and Vivien Raymond.
\newblock {Parameter estimation with a spinning multimode waveform model}.
\newblock {\em Phys. Rev. D}, 101(10):103004, 2020.

\bibitem{Shaik:2019dym}
Feroz~H. Shaik, Jacob Lange, Scott~E. Field, Richard O'Shaughnessy, Vijay
  Varma, Lawrence~E. Kidder, Harald~P. Pfeiffer, and Daniel Wysocki.
\newblock {Impact of subdominant modes on the interpretation of
  gravitational-wave signals from heavy binary black hole systems}.
\newblock {\em Phys. Rev. D}, 101(12):124054, 2020.

\bibitem{LIGOScientific:2020stg}
R.~Abbott et~al.
\newblock {GW190412: Observation of a Binary-Black-Hole Coalescence with
  Asymmetric Masses}.
\newblock {\em Phys. Rev. D}, 102(4):043015, 2020.

\bibitem{LIGOScientific:2020zkf}
R.~Abbott et~al.
\newblock {GW190814: Gravitational Waves from the Coalescence of a 23 Solar
  Mass Black Hole with a 2.6 Solar Mass Compact Object}.
\newblock {\em Astrophys. J. Lett.}, 896(2):L44, 2020.

\bibitem{Colleoni:2020tgc}
Marta Colleoni, Maite Mateu-Lucena, H\'ector Estell\'es, Cecilio
  Garc\'\i{}a-Quir\'os, David Keitel, Geraint Pratten, Antoni Ramos-Buades, and
  Sascha Husa.
\newblock {Towards the routine use of subdominant harmonics in
  gravitational-wave inference: Reanalysis of GW190412 with generation X
  waveform models}.
\newblock {\em Phys. Rev. D}, 103(2):024029, 2021.

\bibitem{Marsat:2020rtl}
Sylvain Marsat, John~G. Baker, and Tito Dal~Canton.
\newblock {Exploring the Bayesian parameter estimation of binary black holes
  with LISA}.
\newblock {\em Phys. Rev. D}, 103(8):083011, 2021.

\bibitem{Baibhav:2020tma}
Vishal Baibhav, Emanuele Berti, and Vitor Cardoso.
\newblock {LISA parameter estimation and source localization with higher
  harmonics of the ringdown}.
\newblock {\em Phys. Rev. D}, 101(8):084053, 2020.

\bibitem{Katz:2021uax}
Michael~L. Katz.
\newblock {Fully automated end-to-end pipeline for massive black hole binary
  signal extraction from LISA data}.
\newblock {\em Phys. Rev. D}, 105(4):044055, 2022.

\bibitem{Krishnendu:2021cyi}
N.~V. Krishnendu and Frank Ohme.
\newblock {Interplay of spin-precession and higher harmonics in the parameter
  estimation of binary black holes}.
\newblock {\em Phys. Rev. D}, 105(6):064012, 2022.

\bibitem{Ng:2022vbz}
Ken K.~Y. Ng et~al.
\newblock {Measuring properties of primordial black hole mergers at
  cosmological distances: Effect of higher order modes in gravitational waves}.
\newblock {\em Phys. Rev. D}, 107(2):024041, 2023.

\bibitem{Pratten:2022kug}
Geraint Pratten, Antoine Klein, Christopher~J. Moore, Hannah Middleton, Nathan
  Steinle, Patricia Schmidt, and Alberto Vecchio.
\newblock {LISA science performance in observations of short-lived signals from
  massive black hole binary coalescences}.
\newblock {\em Phys. Rev. D}, 107(12):123026, 2023.

\bibitem{Pitte:2023ltw}
Chantal Pitte, Quentin Baghi, Sylvain Marsat, Marc Besan\c{c}on, and Antoine
  Petiteau.
\newblock {Detectability of higher harmonics with LISA}.
\newblock {\em Phys. Rev. D}, 108(4):044053, 2023.

\bibitem{Gong:2023ecg}
Yi~Gong, Zhoujian Cao, Junjie Zhao, and Lijing Shao.
\newblock {Including higher harmonics in gravitational-wave parameter
  estimation and cosmological implications for LISA}.
\newblock {\em Phys. Rev. D}, 108(6):064046, 2023.

\bibitem{Hoy:2024aaa}
C~Hoy and L~K Nuttall.
\newblock {bilby in space: Bayesian inference for transient gravitational-wave
  signals observed with LISA}.
\newblock {\em Monthly Notices of the Royal Astronomical Society},
  529(3):3052--3059, 03 2024.

\bibitem{Canizares:2014fya}
Priscilla Canizares, Scott~E. Field, Jonathan Gair, Vivien Raymond, Rory Smith,
  and Manuel Tiglio.
\newblock {Accelerated gravitational-wave parameter estimation with reduced
  order modeling}.
\newblock {\em Phys. Rev. Lett.}, 114(7):071104, 2015.

\bibitem{Vinciguerra:2017ngf}
Serena Vinciguerra, John Veitch, and Ilya Mandel.
\newblock {Accelerating gravitational wave parameter estimation with multi-band
  template interpolation}.
\newblock {\em Class. Quant. Grav.}, 34(11):115006, 2017.

\bibitem{Wysocki:2019grj}
D.~Wysocki, R.~O'Shaughnessy, Jacob Lange, and Yao-Lung~L. Fang.
\newblock {Accelerating parameter inference with graphics processing units}.
\newblock {\em Phys. Rev. D}, 99(8):084026, 2019.

\bibitem{Morisaki:2020oqk}
Soichiro Morisaki and Vivien Raymond.
\newblock {Rapid Parameter Estimation of Gravitational Waves from Binary
  Neutron Star Coalescence using Focused Reduced Order Quadrature}.
\newblock {\em Phys. Rev. D}, 102(10):104020, 2020.

\bibitem{Qi:2020lfr}
Hong Qi and Vivien Raymond.
\newblock {Python-based reduced order quadrature building code for fast
  gravitational wave inference}.
\newblock {\em Phys. Rev. D}, 104(6):063031, 2021.

\bibitem{Morisaki:2021ngj}
Soichiro Morisaki.
\newblock {Accelerating parameter estimation of gravitational waves from
  compact binary coalescence using adaptive frequency resolutions}.
\newblock {\em Phys. Rev. D}, 104(4):044062, 2021.

\bibitem{Williams:2021qyt}
Michael~J. Williams, John Veitch, and Chris Messenger.
\newblock {Nested sampling with normalizing flows for gravitational-wave
  inference}.
\newblock {\em Phys. Rev. D}, 103(10):103006, 2021.

\bibitem{Hoy:2022tst}
Charlie Hoy.
\newblock {Accelerating multimodel Bayesian inference, model selection, and
  systematic studies for gravitational wave astronomy}.
\newblock {\em Phys. Rev. D}, 106(8):083003, 2022.

\bibitem{Pathak:2022iar}
Lalit Pathak, Amit Reza, and Anand~S. Sengupta.
\newblock {Fast likelihood evaluation using meshfree approximations for
  reconstructing compact binary sources}.
\newblock {\em Phys. Rev. D}, 108(6):064055, 2023.

\bibitem{Rose:2022axr}
Caitlin~A. Rose, Vinaya Valsan, Patrick~R. Brady, Sinead Walsh, and Chris
  Pankow.
\newblock {Supplementing rapid Bayesian parameter estimation schemes with
  adaptive grids}.
\newblock {\em arXiv e-prints}, page arXiv:2201.05263, 1 2022.

\bibitem{Dax:2022pxd}
Maximilian Dax, Stephen~R. Green, Jonathan Gair, Michael P\"urrer, Jonas
  Wildberger, Jakob~H. Macke, Alessandra Buonanno, and Bernhard Sch\"olkopf.
\newblock {Neural Importance Sampling for Rapid and Reliable Gravitational-Wave
  Inference}.
\newblock {\em Phys. Rev. Lett.}, 130(17):171403, 2023.

\bibitem{Lee:2022jpn}
Eunsub Lee, Soichiro Morisaki, and Hideyuki Tagoshi.
\newblock {Mass-spin reparametrization for a rapid parameter estimation of
  inspiral gravitational-wave signals}.
\newblock {\em Phys. Rev. D}, 105(12):124057, 2022.

\bibitem{Roulet:2022kot}
Javier Roulet, Seth Olsen, Jonathan Mushkin, Tousif Islam, Tejaswi Venumadhav,
  Barak Zackay, and Matias Zaldarriaga.
\newblock {Removing degeneracy and multimodality in gravitational wave source
  parameters}.
\newblock {\em Phys. Rev. D}, 106(12):123015, 2022.

\bibitem{Yelikar:2023mwg}
A.~B. Yelikar, V.~Delfavero, and R.~O'Shaughnessy.
\newblock {Low-latency parameter inference enabled by a Gaussian likelihood
  approximation for RIFT}.
\newblock {\em arXiv e-prints}, page arXiv:2301.01337, 1 2023.

\bibitem{Morras:2023pug}
Gonzalo Morras, Jose Francisco~Nuno Siles, and Juan Garcia-Bellido.
\newblock {Efficient reduced order quadrature construction algorithms for fast
  gravitational wave inference}.
\newblock {\em Phys. Rev. D}, 108(12):123025, 2023.

\bibitem{Pathak:2023ixb}
Lalit Pathak, Sanket Munishwar, Amit Reza, and Anand~S. Sengupta.
\newblock {Prompt sky localization of compact binary sources using a meshfree
  approximation}.
\newblock {\em Phys. Rev. D}, 109(2):024053, 2024.

\bibitem{Morisaki:2023kuq}
Soichiro Morisaki, Rory Smith, Leo Tsukada, Surabhi Sachdev, Simon Stevenson,
  Colm Talbot, and Aaron Zimmerman.
\newblock {Rapid localization and inference on compact binary coalescences with
  the Advanced LIGO-Virgo-KAGRA gravitational-wave detector network}.
\newblock {\em Phys. Rev. D}, 108(12):123040, 2023.

\bibitem{Williams:2023ppp}
Michael~J. Williams, John Veitch, and Chris Messenger.
\newblock {Importance nested sampling with normalising flows}.
\newblock {\em Mach. Learn. Sci. Tech.}, 4(3):035011, 2023.

\bibitem{Wong:2023lgb}
Kaze W.~K. Wong, Maximiliano Isi, and Thomas D.~P. Edwards.
\newblock {Fast Gravitational-wave Parameter Estimation without Compromises}.
\newblock {\em Astrophys. J.}, 958(2):129, 2023.

\bibitem{Tiwari:2023mzf}
Vaibhav Tiwari, Charlie Hoy, Stephen Fairhurst, and Duncan MacLeod.
\newblock {Fast non-Markovian sampler for estimating gravitational-wave
  posteriors}.
\newblock {\em Phys. Rev. D}, 108(2):023001, 2023.

\bibitem{Fairhurst:2023idl}
Stephen Fairhurst, Charlie Hoy, Rhys Green, Cameron Mills, and Samantha~A.
  Usman.
\newblock {Simple parameter estimation using observable features of
  gravitational-wave signals}.
\newblock {\em Phys. Rev. D}, 108(8):082006, 2023.

\bibitem{Tiwari:2024qzr}
Vaibhav Tiwari.
\newblock {Varaha: A promising sampler for obtaining gravitational wave
  posteriors}.
\newblock {\em arXiv e-prints}, page arXiv:2405.16568, 5 2024.

\bibitem{Wouters:2024oxj}
Thibeau Wouters, Peter T.~H. Pang, Tim Dietrich, and Chris Van Den~Broeck.
\newblock {Robust parameter estimation within minutes on gravitational wave
  signals from binary neutron star inspirals}.
\newblock {\em arXiv e-prints}, page arXiv:2404.11397, 4 2024.

\bibitem{Vilchez:2024qnw}
Iv\'an~Mart\'\i{}n V\'\i{}lchez and Carlos~F. Sopuerta.
\newblock {Efficient Massive Black Hole Binary parameter estimation for LISA
  using Sequential Neural Likelihood}.
\newblock {\em arXiv e-prints}, page arXiv:2406.00565, 6 2024.

\bibitem{Cornish:2010kf}
Neil~J. Cornish.
\newblock {Fast Fisher Matrices and Lazy Likelihoods}.
\newblock {\em arXiv e-prints}, page arXiv:1007.4820, 7 2010.

\bibitem{Zackay:2018qdy}
Barak Zackay, Liang Dai, and Tejaswi Venumadhav.
\newblock {Relative Binning and Fast Likelihood Evaluation for Gravitational
  Wave Parameter Estimation}.
\newblock {\em arXiv e-prints}, page arXiv:1806.08792, 6 2018.

\bibitem{Cornish:2021lje}
Neil~J. Cornish.
\newblock {Heterodyned likelihood for rapid gravitational wave parameter
  inference}.
\newblock {\em Phys. Rev. D}, 104(10):104054, 2021.

\bibitem{Krishna:2023bug}
Kruthi Krishna, Aditya Vijaykumar, Apratim Ganguly, Colm Talbot, Sylvia
  Biscoveanu, Richard~N. George, Natalie Williams, and Aaron Zimmerman.
\newblock {Accelerated parameter estimation in Bilby with relative binning}.
\newblock {\em arXiv e-prints}, page arXiv:2312.06009, 12 2023.

\bibitem{Weaving:2023fji}
Connor~R. Weaving, Laura~K. Nuttall, Ian~W. Harry, Shichao Wu, and Alexander
  Nitz.
\newblock {Adapting the PyCBC pipeline to find and infer the properties of
  gravitational waves from massive black hole binaries in LISA}.
\newblock {\em Class. Quant. Grav.}, 41(2):025006, 2024.

\bibitem{Marsat:2018oam}
Sylvain Marsat and John~G. Baker.
\newblock {Fourier-domain modulations and delays of gravitational-wave
  signals}.
\newblock {\em arXiv e-prints}, page arXiv:1806.10734, 6 2018.

\bibitem{Littenberg:2023xpl}
Tyson~B. Littenberg and Neil~J. Cornish.
\newblock {Prototype global analysis of LISA data with multiple source types}.
\newblock {\em Phys. Rev. D}, 107(6):063004, 2023.

\bibitem{Klein:2022rbf}
Antoine Klein et~al.
\newblock {The last three years: multiband gravitational-wave observations of
  stellar-mass binary black holes}.
\newblock {\em arXiv e-prints}, page arXiv:2204.03423, 4 2022.

\bibitem{Liang:2024new}
Bo~Liang, Minghui Du, He~Wang, Yuxiang Xu, Chang Liu, Xiaotong Wei, Peng Xu,
  Li-e Qiang, and Ziren Luo.
\newblock {Rapid Parameter Estimation for Merging Massive Black Hole Binaries
  Using ODE-Based Generative Models}.
\newblock {\em arXiv e-prints}, page arXiv:2407.07125, 7 2024.

\bibitem{whittle1951hypothesis}
Peter Whittle.
\newblock Hypothesis testing in time series analysis.
\newblock {\em Uppsala: Almqvist \& Wiksells Boktryckeri AB, Oxford, England},
  1951.

\bibitem{Romano:2016dpx}
Joseph~D. Romano and Neil~J. Cornish.
\newblock {Detection methods for stochastic gravitational-wave backgrounds: a
  unified treatment}.
\newblock {\em Living Rev. Rel.}, 20(1):2, 2017.

\bibitem{Thrane:2018qnx}
Eric Thrane and Colm Talbot.
\newblock {An introduction to Bayesian inference in gravitational-wave
  astronomy: parameter estimation, model selection, and hierarchical models}.
\newblock {\em Publ. Astron. Soc. Austral.}, 36:e010, 2019.
\newblock [Erratum: Publ.Astron.Soc.Austral. 37, e036 (2020)].

\bibitem{metropolis1949monte}
Nicholas Metropolis and Stanislaw Ulam.
\newblock The monte carlo method.
\newblock {\em Journal of the American statistical association},
  44(247):335--341, 1949.

\bibitem{Skilling2004}
John Skilling.
\newblock Nested sampling.
\newblock In {\em {AIP} Conference Proceedings}. {AIP}, 2004.

\bibitem{Skilling:2006gxv}
John Skilling.
\newblock {Nested sampling for general Bayesian computation}.
\newblock {\em Bayesian Analysis}, 1(4):833--859, 2006.

\bibitem{Veitch:2014wba}
J.~Veitch et~al.
\newblock {Parameter estimation for compact binaries with ground-based
  gravitational-wave observations using the LALInference software library}.
\newblock {\em Phys. Rev. D}, 91(4):042003, 2015.

\bibitem{Ashton:2018jfp}
Gregory Ashton et~al.
\newblock {BILBY: A user-friendly Bayesian inference library for
  gravitational-wave astronomy}.
\newblock {\em Astrophys. J. Suppl.}, 241(2):27, 2019.

\bibitem{Romero-Shaw:2020owr}
I.~M. Romero-Shaw et~al.
\newblock {Bayesian inference for compact binary coalescences with bilby:
  validation and application to the first LIGO\textendash{}Virgo
  gravitational-wave transient catalogue}.
\newblock {\em Mon. Not. Roy. Astron. Soc.}, 499(3):3295--3319, 2020.

\bibitem{Biwer:2018osg}
C.~M. Biwer, Collin~D. Capano, Soumi De, Miriam Cabero, Duncan~A. Brown,
  Alexander~H. Nitz, and V.~Raymond.
\newblock {PyCBC Inference: A Python-based parameter estimation toolkit for
  compact binary coalescence signals}.
\newblock {\em Publ. Astron. Soc. Pac.}, 131(996):024503, 2019.

\bibitem{Pankow:2015cra}
C.~Pankow, P.~Brady, E.~Ochsner, and R.~O'Shaughnessy.
\newblock {Novel scheme for rapid parallel parameter estimation of
  gravitational waves from compact binary coalescences}.
\newblock {\em Phys. Rev. D}, 92(2):023002, 2015.

\bibitem{Lange:2018pyp}
Jacob Lange, Richard O'Shaughnessy, and Monica Rizzo.
\newblock {Rapid and accurate parameter inference for coalescing, precessing
  compact binaries}.
\newblock {\em arXiv e-prints}, page arXiv:1805.10457, 5 2018.

\bibitem{Delaunoy:2020zcu}
Arnaud Delaunoy, Antoine Wehenkel, Tanja Hinderer, Samaya Nissanke, Christoph
  Weniger, Andrew~R. Williamson, and Gilles Louppe.
\newblock {Lightning-Fast Gravitational Wave Parameter Inference through Neural
  Amortization}.
\newblock {\em arXiv e-prints}, page arXiv:2010.12931, 10 2020.

\bibitem{Green:2020hst}
Stephen~R. Green, Christine Simpson, and Jonathan Gair.
\newblock {Gravitational-wave parameter estimation with autoregressive neural
  network flows}.
\newblock {\em Phys. Rev. D}, 102(10):104057, 2020.

\bibitem{Chua:2019wwt}
Alvin J.~K. Chua and Michele Vallisneri.
\newblock {Learning Bayesian posteriors with neural networks for
  gravitational-wave inference}.
\newblock {\em Phys. Rev. Lett.}, 124(4):041102, 2020.

\bibitem{Green:2020dnx}
Stephen~R. Green and Jonathan Gair.
\newblock {Complete parameter inference for GW150914 using deep learning}.
\newblock {\em Mach. Learn. Sci. Tech.}, 2(3):03LT01, 2021.

\bibitem{Dax:2021tsq}
Maximilian Dax, Stephen~R. Green, Jonathan Gair, Jakob~H. Macke, Alessandra
  Buonanno, and Bernhard Sch\"olkopf.
\newblock {Real-Time Gravitational Wave Science with Neural Posterior
  Estimation}.
\newblock {\em Phys. Rev. Lett.}, 127(24):241103, 2021.

\bibitem{Gabbard:2019rde}
Hunter Gabbard, Chris Messenger, Ik~Siong Heng, Francesco Tonolini, and
  Roderick Murray-Smith.
\newblock {Bayesian parameter estimation using conditional variational
  autoencoders for gravitational-wave astronomy}.
\newblock {\em Nature Phys.}, 18(1):112--117, 2022.

\bibitem{Owen:1995tm}
Benjamin~J. Owen.
\newblock {Search templates for gravitational waves from inspiraling binaries:
  Choice of template spacing}.
\newblock {\em Phys. Rev. D}, 53:6749--6761, 1996.

\bibitem{Baird:2012cu}
Emily Baird, Stephen Fairhurst, Mark Hannam, and Patricia Murphy.
\newblock {Degeneracy between mass and spin in black-hole-binary waveforms}.
\newblock {\em Phys. Rev. D}, 87(2):024035, 2013.

\bibitem{Owen:1998dk}
Benjamin~J. Owen and B.~S. Sathyaprakash.
\newblock {Matched filtering of gravitational waves from inspiraling compact
  binaries: Computational cost and template placement}.
\newblock {\em Phys. Rev. D}, 60:022002, 1999.

\bibitem{Cutler:1994ys}
Curt Cutler and Eanna~E. Flanagan.
\newblock {Gravitational waves from merging compact binaries: How accurately
  can one extract the binary's parameters from the inspiral wave form?}
\newblock {\em Phys. Rev. D}, 49:2658--2697, 1994.

\bibitem{Ohme:2013nsa}
Frank Ohme, Alex~B. Nielsen, Drew Keppel, and Andrew Lundgren.
\newblock {Statistical and systematic errors for gravitational-wave inspiral
  signals: A principal component analysis}.
\newblock {\em Phys. Rev. D}, 88(4):042002, 2013.

\bibitem{Farr:2014qka}
Benjamin Farr, Evan Ochsner, Will~M. Farr, and Richard O'Shaughnessy.
\newblock {A more effective coordinate system for parameter estimation of
  precessing compact binaries from gravitational waves}.
\newblock {\em Phys. Rev. D}, 90(2):024018, 2014.

\bibitem{Usman:2015kfa}
Samantha~A. Usman et~al.
\newblock {The PyCBC search for gravitational waves from compact binary
  coalescence}.
\newblock {\em Class. Quant. Grav.}, 33(21):215004, 2016.

\bibitem{pycbc-software}
Alex Nitz, Ian Harry, Duncan Brown, Christopher~M. Biwer, Josh Willis, Tito~Dal
  Canton, Collin Capano, Larne Pekowsky, Thomas Dent, Andrew~R. Williamson,
  et~al.
\newblock gwastro/pycbc: Pycbc release v1.15.2.
\newblock Zenodo, December 2019.

\bibitem{Tinto:1999yr}
Massimo Tinto and J.~W. Armstrong.
\newblock {Cancellation of laser noise in an unequal-arm interferometer
  detector of gravitational radiation}.
\newblock {\em Phys. Rev. D}, 59:102003, 1999.

\bibitem{armstrong1999time}
JW~Armstrong, FB~Estabrook, and Massimo Tinto.
\newblock Time-delay interferometry for space-based gravitational wave
  searches.
\newblock {\em The Astrophysical Journal}, 527(2):814, 1999.

\bibitem{Estabrook:2000ef}
F.~B. Estabrook, Massimo Tinto, and J.~W. Armstrong.
\newblock {Time delay analysis of LISA gravitational wave data: Elimination of
  spacecraft motion effects}.
\newblock {\em Phys. Rev. D}, 62:042002, 2000.

\bibitem{Vallisneri:2005ji}
Michele Vallisneri.
\newblock {Geometric time delay interferometry}.
\newblock {\em Phys. Rev. D}, 72:042003, 2005.
\newblock [Erratum: Phys.Rev.D 76, 109903 (2007)].

\bibitem{Tinto:2020fcc}
Massimo Tinto and Sanjeev~V. Dhurandhar.
\newblock {Time-delay interferometry}.
\newblock {\em Living Rev. Rel.}, 24(1):1, 2021.

\bibitem{tinto2023second}
Massimo Tinto, Sanjeev Dhurandhar, and Dishari Malakar.
\newblock Second-generation time-delay interferometry.
\newblock {\em Physical Review D}, 107(8):082001, 2023.

\bibitem{Prince:2002hp}
Thomas~A. Prince, Massimo Tinto, Shane~L. Larson, and J.~W. Armstrong.
\newblock {The LISA optimal sensitivity}.
\newblock {\em Phys. Rev. D}, 66:122002, 2002.

\bibitem{Katz:2020hku}
Michael~L. Katz, Sylvain Marsat, Alvin J.~K. Chua, Stanislav Babak, and
  Shane~L. Larson.
\newblock {GPU-accelerated massive black hole binary parameter estimation with
  LISA}.
\newblock {\em Phys. Rev. D}, 102(2):023033, 2020.

\bibitem{michael_katz_2021_5730688}
Michael Katz.
\newblock mikekatz04/bbhx: First official public release, November 2021.

\bibitem{London:2017bcn}
Lionel London, Sebastian Khan, Edward Fauchon-Jones, Cecilio García, Mark
  Hannam, Sascha Husa, Xisco Jiménez-Forteza, Chinmay Kalaghatgi, Frank Ohme,
  and Francesco Pannarale.
\newblock {First higher-multipole model of gravitational waves from spinning
  and coalescing black-hole binaries}.
\newblock {\em PhRvL.}, 120(16):161102, 2018.

\bibitem{Vecchio:2003tn}
Alberto Vecchio.
\newblock {LISA observations of rapidly spinning massive black hole binary
  systems}.
\newblock {\em Phys. Rev. D}, 70:042001, 2004.

\bibitem{Berti:2004bd}
Emanuele Berti, Alessandra Buonanno, and Clifford~M. Will.
\newblock {Estimating spinning binary parameters and testing alternative
  theories of gravity with LISA}.
\newblock {\em Phys. Rev. D}, 71:084025, 2005.

\bibitem{Lang:2006bsg}
Ryan~N. Lang and Scott~A. Hughes.
\newblock {Measuring coalescing massive binary black holes with gravitational
  waves: The Impact of spin-induced precession}.
\newblock {\em Phys. Rev. D}, 74:122001, 2006.
\newblock [Erratum: Phys.Rev.D 75, 089902 (2007), Erratum: Phys.Rev.D 77,
  109901 (2008)].

\bibitem{Arun:2008zn}
K.~G. Arun et~al.
\newblock {Massive Black Hole Binary Inspirals: Results from the LISA Parameter
  Estimation Taskforce}.
\newblock {\em Class. Quant. Grav.}, 26:094027, 2009.

\bibitem{Speagle:2019ivv}
Joshua~S. Speagle.
\newblock {dynesty: a dynamic nested sampling package for estimating Bayesian
  posteriors and evidences}.
\newblock {\em Mon. Not. Roy. Astron. Soc.}, 493(3):3132--3158, 2020.

\bibitem{LIGOScientific:2016dsl}
B.~P. Abbott et~al.
\newblock {Binary Black Hole Mergers in the first Advanced LIGO Observing Run}.
\newblock {\em Phys. Rev. X}, 6(4):041015, 2016.
\newblock [Erratum: Phys.Rev.X 8, 039903 (2018)].

\bibitem{Leaver:1985ax}
E.~W. Leaver.
\newblock {An Analytic representation for the quasi normal modes of Kerr black
  holes}.
\newblock {\em Proc. Roy. Soc. Lond. A}, 402:285--298, 1985.

\bibitem{Berti:2005ys}
Emanuele Berti, Vitor Cardoso, and Clifford~M. Will.
\newblock {On gravitational-wave spectroscopy of massive black holes with the
  space interferometer LISA}.
\newblock {\em Phys. Rev. D}, 73:064030, 2006.

\bibitem{Roy:2019phx}
Soumen Roy, Anand~S. Sengupta, and K.~G. Arun.
\newblock {Unveiling the spectrum of inspiralling binary black holes}.
\newblock {\em Phys. Rev. D}, 103(6):064012, 2021.

\bibitem{Blanchet:2013haa}
Luc Blanchet.
\newblock {Gravitational Radiation from Post-Newtonian Sources and Inspiralling
  Compact Binaries}.
\newblock {\em Living Rev. Rel.}, 17:2, 2014.

\bibitem{Varma:2018mmi}
Vijay Varma, Scott~E. Field, Mark~A. Scheel, Jonathan Blackman, Lawrence~E.
  Kidder, and Harald~P. Pfeiffer.
\newblock {Surrogate model of hybridized numerical relativity binary black hole
  waveforms}.
\newblock {\em Phys. Rev. D}, 99(6):064045, 2019.

\bibitem{Cotesta:2018fcv}
Roberto Cotesta, Alessandra Buonanno, Alejandro Boh\'e, Andrea Taracchini, Ian
  Hinder, and Serguei Ossokine.
\newblock {Enriching the Symphony of Gravitational Waves from Binary Black
  Holes by Tuning Higher Harmonics}.
\newblock {\em Phys. Rev. D}, 98(8):084028, 2018.

\bibitem{Pompili:2023tna}
Lorenzo Pompili et~al.
\newblock {Laying the foundation of the effective-one-body waveform models
  SEOBNRv5: Improved accuracy and efficiency for spinning nonprecessing binary
  black holes}.
\newblock {\em Phys. Rev. D}, 108(12):124035, 2023.

\bibitem{Fairhurst:2023beb}
Stephen Fairhurst, Cameron Mills, Monica Colpi, Raffaella Schneider, Alberto
  Sesana, Alessandro Trinca, and Rosa Valiante.
\newblock {Identifying heavy stellar black holes at cosmological distances with
  next-generation gravitational-wave observatories}.
\newblock {\em Mon. Not. Roy. Astron. Soc.}, 529(3):2116--2130, 2024.

\bibitem{Leslie:2021ssu}
Nathaniel Leslie, Liang Dai, and Geraint Pratten.
\newblock {Mode-by-mode relative binning: Fast likelihood estimation for
  gravitational waveforms with spin-orbit precession and multiple harmonics}.
\newblock {\em Phys. Rev. D}, 104(12):123030, 2021.

\bibitem{Weaving:2024the}
Connor Weaving.
\newblock {LISA data analysis with massive black hole binary mergers, PhD
  thesis}.
\newblock \emph{in preparation}, 2024.

\bibitem{Harry:2009ea}
Ian~W. Harry, Bruce Allen, and B.~S. Sathyaprakash.
\newblock {A Stochastic template placement algorithm for gravitational wave
  data analysis}.
\newblock {\em Phys. Rev. D}, 80:104014, 2009.

\bibitem{Ajith:2012mn}
P.~Ajith, N.~Fotopoulos, S.~Privitera, A.~Neunzert, and A.~J. Weinstein.
\newblock {Effectual template bank for the detection of gravitational waves
  from inspiralling compact binaries with generic spins}.
\newblock {\em Phys. Rev. D}, 89(8):084041, 2014.

\bibitem{Sachdev:2019vvd}
Surabhi Sachdev et~al.
\newblock {The GstLAL Search Analysis Methods for Compact Binary Mergers in
  Advanced LIGO's Second and Advanced Virgo's First Observing Runs}.
\newblock {\em arXiv e-prints}, page arXiv:1901.08580, 1 2019.

\bibitem{Andres:2021vew}
Nicolas Andres et~al.
\newblock {Assessing the compact-binary merger candidates reported by the MBTA
  pipeline in the LIGO\textendash{}Virgo O3 run: probability of astrophysical
  origin, classification, and associated uncertainties}.
\newblock {\em Class. Quant. Grav.}, 39(5):055002, 2022.

\bibitem{Harry:2017weg}
Ian Harry, Juan Calder\'on~Bustillo, and Alex Nitz.
\newblock {Searching for the full symphony of black hole binary mergers}.
\newblock {\em Phys. Rev. D}, 97(2):023004, 2018.

\bibitem{Chandra:2022ixv}
Koustav Chandra, Juan Calder\'on~Bustillo, Archana Pai, and I.~W. Harry.
\newblock {First gravitational-wave search for intermediate-mass black hole
  mergers with higher-order harmonics}.
\newblock {\em Phys. Rev. D}, 106(12):123003, 2022.

\bibitem{Wadekar:2023kym}
Digvijay Wadekar, Tejaswi Venumadhav, Ajit~Kumar Mehta, Javier Roulet, Seth
  Olsen, Jonathan Mushkin, Barak Zackay, and Matias Zaldarriaga.
\newblock {A new approach to template banks of gravitational waves with higher
  harmonics: reducing matched-filtering cost by over an order of magnitude}.
\newblock {\em arXiv e-prints}, page arXiv:2310.15233, 10 2023.

\bibitem{le_jeune_2022_7132178}
Maude Le~Jeune and Stanislav Babak.
\newblock Lisa data challenge sangria (ldc2a), October 2022.

\bibitem{Hannam:2021pit}
Mark Hannam et~al.
\newblock {General-relativistic precession in a black-hole binary}.
\newblock {\em Nature}, 610(7933):652--655, 2022.

\bibitem{Sesana:2021jfh}
Alberto Sesana.
\newblock {Black Hole Science With the Laser Interferometer Space Antenna}.
\newblock {\em Front. Astron. Space Sci.}, 8:601646, 2021.

\bibitem{Katz:2024oqg}
Michael~L. Katz, Nikolaos Karnesis, Natalia Korsakova, Jonathan~R. Gair, and
  Nikolaos Stergioulas.
\newblock {An efficient GPU-accelerated multi-source global fit pipeline for
  LISA data analysis}.
\newblock {\em arXiv e-prints}, page arXiv:2405.04690, 5 2024.

\bibitem{Strub:2024kbe}
Stefan~H. Strub, Luigi Ferraioli, C\'edric Schmelzbach, Simon~C. St\"ahler, and
  Domenico Giardini.
\newblock {Global analysis of LISA data with Galactic binaries and massive
  black hole binaries}.
\newblock {\em Phys. Rev. D}, 110(2):024005, 2024.

\bibitem{LISAConsortiumWaveformWorkingGroup:2023arg}
Niaesh Afshordi et~al.
\newblock {Waveform Modelling for the Laser Interferometer Space Antenna}.
\newblock {\em arXiv e-prints}, page arXiv:2311.01300, 11 2023.

\bibitem{Apostolatos:1994mx}
Theocharis~A. Apostolatos, Curt Cutler, Gerald~J. Sussman, and Kip~S. Thorne.
\newblock {Spin induced orbital precession and its modulation of the
  gravitational wave forms from merging binaries}.
\newblock {\em Phys. Rev. D}, 49:6274--6297, 1994.

\bibitem{harris2020array}
Charles~R. Harris et~al.
\newblock {Array programming with NumPy}.
\newblock {\em Nature}, 585(7825):357--362, 2020.

\bibitem{2020SciPy-NMeth}
Pauli Virtanen, Ralf Gommers, Travis~E. Oliphant, Matt Haberland, Tyler Reddy,
  David Cournapeau, Evgeni Burovski, Pearu Peterson, Warren Weckesser, Jonathan
  Bright, St{\'e}fan~J. {van der Walt}, Matthew Brett, Joshua Wilson, K.~Jarrod
  Millman, Nikolay Mayorov, Andrew R.~J. Nelson, Eric Jones, Robert Kern, Eric
  Larson, C~J Carey, {\.I}lhan Polat, Yu~Feng, Eric~W. Moore, Jake
  {VanderPlas}, Denis Laxalde, Josef Perktold, Robert Cimrman, Ian Henriksen,
  E.~A. Quintero, Charles~R. Harris, Anne~M. Archibald, Ant{\^o}nio~H. Ribeiro,
  Fabian Pedregosa, Paul {van Mulbregt}, and {SciPy 1.0 Contributors}.
\newblock {{SciPy} 1.0: Fundamental Algorithms for Scientific Computing in
  Python}.
\newblock {\em Nature Methods}, 17:261--272, 2020.

\bibitem{Hoy:2020vys}
Charlie Hoy and Vivien Raymond.
\newblock {PESummary: the code agnostic Parameter Estimation Summary page
  builder}.
\newblock {\em SoftwareX}, 15:100765, 2021.

\bibitem{2007CSE.....9...90H}
J.~D. {Hunter}.
\newblock {Matplotlib: A 2D Graphics Environment}.
\newblock {\em CSE}, 9:90--95, May 2007.

\end{thebibliography}

\end{document}